\documentclass[aps,prb,amsmath,amsfonts,amssymb,showpacs,superscriptaddress,floatfix,twocolumn]{revtex4-1}
\usepackage{times}
\usepackage{epsfig}
\usepackage{bm}
\usepackage{color}
\newcommand{\cR}{c_{\text{R}}}           
\newcommand{\cS}{c_{\text{S}}}           
\newcommand{\cL}{c_{\text{L}}}           
\newcommand{\vi}{v_{\text{i}}}           
\newcommand{\vbar}{\overline{v}}         
\newcommand{\ai}{a_{\text{i}}}           
\newcommand{\dd}{\mathrm{d}}             
\newcommand{\ii}{\mathrm{i}}             
\newcommand{\siga}{\sigma_{\text{a}}}    
\newcommand{\sigth}{\sigma_{\text{th}}}  
\newcommand{\braket}[2]{\left\langle #1|#2\right\rangle} 
\renewcommand{\Im}{\mathop{\mathrm{Im}}} 
\renewcommand{\Re}{\mathop{\mathrm{Re}}} 

\begin{document}
\title{Equation of motion and subsonic-transonic transitions of rectilinear edge dislocations:\\
 A collective-variable approach}
\author{Yves-Patrick \surname{Pellegrini}}
\email{yves-patrick.pellegrini@cea.fr}
\affiliation{CEA, DAM, DIF, F-91297 Arpajon, France.}
\date{Received 18 July 2013; revised manuscript received 4 July 2014; published 29 August 2014}
\begin{abstract}
A theoretical framework is proposed to derive a dynamic equation motion for rectilinear dislocations within isotropic continuum elastodynamics. The theory relies on a recent dynamic extension of the Peierls-Nabarro equation, so as to account for core-width generalized stacking-fault energy effects. The degrees of freedom of the solution of the latter equation are reduced by means of the collective-variable method, well-known in soliton theory, which we reformulate in a way suitable to the problem at hand. Through these means, two coupled governing equations for the dislocation position and core width are obtained, which are combined into one single complex-valued equation of motion, of compact form. The latter equation embodies the history dependence of dislocation inertia. It is employed to investigate the motion of an edge dislocation under uniform time-dependent loading, with focus on the subsonic/transonic transition. Except in the steady-state supersonic range of velocities\----\-which the equation does not address\----\-our results are in good agreement with atomistic simulations on tungsten. In particular, we provide an explanation for the transition, showing that it is governed by a loading-dependent \emph{dynamic} critical stress. The transition has the character of a delayed bifurcation. Moreover, various quantitative predictions are made, that could be tested in atomistic simulations. Overall, this work demonstrates the crucial role played by core-width variations in dynamic dislocation motion.
\end{abstract}
\hbox{ArXiv version of: {\small PHYSICAL REVIEW B} \textbf{90}, 054120 (2014) [18 pages].}\hfill
\maketitle
\section{Introduction}
\label{sec:Introduction}
In the last decade, high-velocity dislocation motion in crystals has been the subject of many two-dimensional (2D) studies by molecular dynamics,\cite{GUMB99,LI02,MORD03,VAND04,OLMS05,MARI06,MORD06,TSUZ08,JING08} or via direct measurements in a plasma crystal slab.\cite{NOSE07,MORF09} From these studies, a wealth of data has been collected, either in the form of time-velocity curves,\cite{OLMS05} or of terminal velocity-versus-applied stress plots.\cite{GUMB99,JING08} These data illustrate the wide variety of behaviors presented by dislocations subjected to dynamic loadings in the 2D problem. Among some intriguing effects, the subsonic/transonic transition of an edge dislocation\cite{GUMB99,JING08} deserves special attention, in view of its complicated dependence on the external loading. For instance, the long-time asymptotic velocity state of the dislocation depends on the loading being applied in one, or in two steps, which no theory has yet reproduced. During the transition the dislocation core undergoes drastic size variations.\cite{GUMB99,JING08} Such dynamic effects are expected to be important to high-strain rate deformation processes, and could notably modify the well-established interaction mechanisms between dislocations in quasistatic deformation.\cite{PILL06} So far, discussion of the transonic/subsonic transition has mainly been based either on the atomistic simulations, or on steady-state models.\cite{WEER69b,WEER69a,ROSA01} It is natural, however, to expect that more complete understanding of the phenomenon will arise from a suitable dynamic equation of motion (EoM) of dislocations---yet to be found.

Apart from their intrinsic theoretical interest, the above issues must clearly be elucidated in the recently adv\-oc\-a\-ted perspective of employing non-quasistatic dislocation-dynamics methods to study fast deformation processes in metals at the mesoscopic level, accounting for elastodynamic wave propagation.\cite{GURR13} Traditionally, approaches to dislocation motion, mainly relevant to the low-velocity phonon-drag-controlled regime,\cite{NADG88} rely on a simple overdamped mobility law and include inertia (when needed) by means of a Newtonian-like dislocation mass factor.\cite{LUND8586,BITZ05} However, such phenomenological approaches (including relativistic ones\cite{HIRT98}) have been ruled out by phase-field calculations when acceleration is fast, and when the velocity becomes a sizable fraction of the shear wave speed.\cite{PILL07} As has long been recognized,\cite{ESHE53} the key to dislocation inertia resides in the phenomenon of radiation reaction. At nonsupersonic velocities, inertia arises from the finite-width dislocation moving within its own wavefield emitted at every past instant, which results in retarded self-interaction and makes motion history-dependent.\cite{ESHE53,BELT68,PILL07,PELL12} For high strain-rate processes, usual mobility laws are simply not adequate.

Starting with Eshelby,\cite{ESHE53} progresses have been made over the years in computing dynamic self-interactions and fields radiated into the surrounding medium by non-uniformly moving dislocations.\cite{GURR13} The self-force has been extensively discussed by Markenscoff and co-workers under an assumptions of rigid dislocation core.\cite{CLIF81} Based on Eshelby's seminal work (valid at small velocities), the issue of radiative-damping losses has been addressed by Al'shits \textit{et al.},\cite{ALSH71} and a phenomenological EoM aimed at velocities less than the shear wave-speed has been proposed by Pillon \emph{et al.}\cite{PILL07} More recently, these questions were re-examined by the present author, resulting in an extended EoM that, in principle, allows for arbitrary core variations with time.\cite{PELL12} However, the theory still lacks an independent governing equation for the core, and must be completed by an ad-hoc approximation to be of use, which considerably restricts its domain of validity. This calls for a more firmly grounded derivation method, capable of providing the missing equation.

Accordingly, the present work aims at obtaining from general principles an improved EoM, in the framework of isotropic elas\-to\-dyn\-amics, paying\----\-in view of the aforementioned available data\----\-special attention to its predictions as to the so-called transonic regime\cite{WEER67b} and to the subsonic/transonic transition. We restrict our attention to single rectilinear dislocations (two-dimensional problem). In particular, the question of dislocation splitting into partials,\cite{BITZ05} which is important for instance in high strain-rate twinning processes,\cite{MARI04} is left out. Steady supersonic motion is not considered either, for reasons to be clarified. However, \emph{transient} supersonic states will be permitted.

The Peierls model proves a convenient starting point to account for core-related effects, via an input from the generalized stacking-fault (GSF) energy (or $\gamma$-surface) potential, which measures the energy cost of lattice mismatch at the core.\cite{VITE68,SCHO01} In statics, the Peierls-Nabarro (PN) functional equation\cite{PEIE40,NABA47,HIRT82,SCHO05} determines the dislocation core shape from balancing the projections on the glide plane of the (self-) stress generated by the dislocation, and of the pull-back stress that keeps the lattice together. The latter stress derives from the GSF energy. Following Eshelby's ideas,\cite{ESHE53} the PN equation has been generalized by the author to dynamics\cite{PELL10,PELL11} into what is called hereafter the \emph{dynamic Peierls-Nabarro equation} (DPE), to be reviewed in Sec.\ \ref{sec:dpe}. In short, the DPE stems from a reduction to one dimension (the coordinate along the glide plane) of the two-dimensional elastodynamic problem that consists in computing in the surrounding medium the elastodynamic fields emitted by the dislocation, in the spirit of boundary-integral-equation methods,\cite{BONN95}. While the latter problem can be addressed by means of phase-field type numerical methods of solution,\cite{DENO04} the computation of the time-dependent fields outside the slip plane proves superfluous for the EoM, for which only the stress field on the slip plane is needed.\cite{PELL11}

Solving numerically the DPE is an outstanding task. Here, recourse is made to approximations. As the DPE deals with infinitely many degrees of freedom (i.e., the full core shape function),\cite{ESHE53} our first step is to reduce this number by employing a suitable ansatz for the master core shape, centered on the dislocation position $\xi(t)$ at time $t$ and scaled by a time-dependent core width variable $a(t)$. These two quantities stand as collective variables (CVs), for which simpler governing equation of motion are sought. To this purpose we appeal to the systematic collective-variable method of projection,\cite{NINO72,TOMB75,BOES88,BOES90,SCHN00} well-known in soliton theory,\cite{PEYR04} which we reformulate in Sec.\ \ref{sec:collvars} in a way that suits our needs. Further details are provided in Appendix \ref{sec:evoleq}. Applying the method to the DPE, we arrive at the desired EoM for the dislocation in Sec.\ \ref{sec:eom}, in the form of a retarded integro-differential functional equation for $\xi(t)$ and $a(t)$, with history dependence in the latter variables. Rather unexpectedly, the EoM naturally shows up in complex-valued form, of real and imaginary parts the (coupled) governing equation for $\xi$ and $a$, respectively. Its real part turns out to be the previously obtained incomplete EoM,\cite{PELL12} while the imaginary part provides the missing equation for the core width. By construction, the obtained EoM will be seen to reduce in the steady-state limit to Rosakis's Model I,\cite{ROSA01} which describes high-velocity steady motion in the nonsupersonic range. As a byproduct, we retrieve the kinetic relations of the latter model in a generic form, independent of the dislocation character. The physical content of the EoM is further explored in Sec.\ \ref{sec:local}. Specializing the EoM to an edge dislocation, and solving it by means of a specially-devised numerical algorithm (Appendix \ref{sec:nummeth}), we carry out in Sec.\ \ref{sec:numres} an in-depth analysis of the subsonic/transonic transition under single-step and double-step loadings, making quantitative comparisons with some atomistic results by Jin \textit{et al}.\cite{JING08} A concluding discussion closes the paper.

Before proceeding, it is worthwhile to point out that in field theories other than elastodynamics, the influence of radiation reaction on the motion of charged particles is of considerable interest as well (for a dislocation, the ``charge" is the Burgers vector). For instance, in classical electrodynamics, the long-standing issue of finding a non-pathological equation of motion for an extended electric charge is still attracting attention. In the latter context, for lack of an appropriate force model to bind the charge distribution together, an hypothesis of rigid shape is almost always made, which leads to conceptual difficulties.\cite{MEDI06} By contrast, the EoM for dislocations to be obtained is internally consistent within elastodynamics, and allows one to study dynamic shape variations of the defect.

\section{The dynamic Peierls-Nabarro equation}
\label{sec:dpe}
In a two-dimensional set-up, the DPE\cite{PELL10,PELL11} describes a rectilinear dislocation with a flat core that moves on its glide plane, under the action of a time-dependent applied shear stress $\sigma_a(t)$, assumed uniform on the glide plane.\cite{NOTE1}

Hereafter, $x$ stands for the coordinate along the direction of motion, and $t$ is the time. Peierls-type models can be viewed as models of the so-called cohesive-zone type\cite{MILL98} that assume nonlinear elasticity in a region of vanishing width surrounding the glide plane, and linear elasticity elsewhere. The nonlinear elastic force-balance law reads
\begin{equation}
\label{eq:nlin}
\sigma_a(t)-f'(\eta(t))=0,
\end{equation}
where $\eta(t)$ represents a uniform relative material displacement (``slip") between both sides of the glide plane, and $-f'(\eta)$ is the ``pullback" force that binds together, under shear deformation, the two atomic planes separated by a distance $d$ that surround the glide plane. This force, of lattice origin, derives from the cohesive potential $f(\eta)$ of periodicity the Burgers vector modulus $b$, often identified with the GSF energy potential.\cite{SCHO01} In this paper, calculations are made with the usual Frenkel sine force
\begin{equation}
\label{eq:pullback}
f'(\eta)=\sigth\sin(2\pi\eta/b),
\end{equation}
where, with $\mu$ the shear modulus,
\begin{eqnarray}
\label{eq:sigth}
\sigth=\max_\eta f'(\eta)=\frac{\mu b}{2\pi d},
\end{eqnarray}
is the theoretical shear stress.\cite{HIRT82} When $\eta$ is small, Eqs.\ (\ref{eq:nlin}) and (\ref{eq:pullback}) reduce to a linear elasticity law $\sigma_a=2\mu\varepsilon$ where $\varepsilon=\eta/(2d)$ represents the elastic shear strain.\cite{HIRT82} We call $\eta^{\rm e}$ the uniform ``background" solution of Eqs.\ (\ref{eq:nlin}) and (\ref{eq:pullback}), namely,
\begin{equation}
\label{eq:etainfty}
\eta^e=\frac{b}{2\pi}\mathop{\rm Arcsin}(\sigma_a/\sigth)\qquad (|\sigma_a|\leq\sigth),
\end{equation}
which describes purely elastic response. It saturates at $\pm b/4$ for $\sigma_a=\pm\sigth$. In presence of a single dislocation of Burgers vector $b$, the slip becomes inhomogeneous. We write it $\eta(x,t)$. It differs from $\eta^{\rm e}(t)$ by the quantity
\begin{equation}
\widetilde{\eta}(x,t)\equiv \eta(x,t)-\eta^{\rm e}(t),
\end{equation}
which characterizes the dislocation and stands as a local Burgers vector component. The field $\widetilde{\eta}$ (or $\eta$) is the unknown of the problem. For a single dislocation coming from $x=-\infty$, boundary conditions are $\widetilde{\eta}(-\infty,t)=b$ and $\widetilde{\eta}(+\infty,t)=0$. We define the (signed) dislocation density as
\begin{equation}
\label{eq:ddens}
\rho=\partial\widetilde{\eta}/\partial x.
\end{equation}
Whereas the usual definition of the density in the Peierls model is $\rho=\partial\eta/\partial x$, both definitions are equivalent here since $\eta^{\rm e}$ is uniform. Equation (\ref{eq:ddens}) introduces a slight modification with respect to previous work,\cite{PELL10,PELL12} and is further commented at the end of this Section.

The DPE that determines $\eta(x,t)$ is a nonlinear integro-differential equation. We write it for convenience as the dynamic equilibrium equation
\begin{equation}
\label{eq:dyncore}
\mathcal{F}(x,t,[\eta])=0,
\end{equation}
where
\begin{equation}
\label{eq:fdef}
\mathcal{F}(x,t,[\eta])=\sigma_\eta(x,t)+\sigma_D(x,t)+\sigma_a(t)-f'\left(\eta(x,t)\right)
\end{equation}
is the total force acting on the dislocation. The term $\sigma_D$ is a phenomenological drag force,\cite{GILM68} written as\cite{WEER69b}
\begin{equation}
\label{eq:dragstress}
\sigma_D(x,t)=-\alpha\frac{\mu}{2\cS}\frac{\partial
\widetilde{\eta}}{\partial t}(x,t),
\end{equation}
where $\cS$ is the shear wave speed, the longitudinal wave speed being written $\cL$ hereafter. The dimensionless friction parameter $\alpha$ embodies various drag mechanisms of non-radiative origin (e.g., Ref.\ \onlinecite{HIRT82} p.\ 209), among which phonon drag is the main contributor at usual temperatures.\cite{NADG88} In order to simplify the writing of Eq.\ (\ref{eq:av}) below and like, $\alpha$ is defined here as \emph{twice the $\alpha$ coefficient of Refs.\ \onlinecite{PELL12,ROSA01}.} As it does not act on $\eta^{\rm e}(t)$, this force can be termed ``viscoplastic".

The term $\sigma_\eta$ is the (retarded) dynamic stress induced by the dislocation.\cite{NOTE2} As will be seen below, it stands for the negative of the inertial self-force on the dislocation. Its expression stems from linear elasticity theory.\cite{PELL10,PELL11,PELL12} Setting $\Delta x=x-x'$ and $\Delta t=t-\tau$, it reads
\begin{equation}
\label{eq:pnstress}
\sigma_\eta(x,t)=-\frac{\mu}{\pi}\int\mathrm{d}\tau\,\mathrm{d}x'\,K(\Delta x,\Delta t)\rho(x',\tau)-\frac{\mu}{2\cS}\frac{\partial\widetilde{\eta}}{\partial t}(x,t),
\end{equation}
where the kernel $K$ depends on the character of the dislocation. It accounts for in-plane wave-propagation effects, \emph{and} for out-of-plane acceleration-radiation (i.e., \textit{Bremsstrahlung}) losses.\cite{NOTE3} The ``local" term in (\ref{eq:pnstress}), proportional to $\partial\widetilde{\eta}/\partial t$, represents another sort of out-of-plane velocity-dependent radiative losses. Apparently first noticed in the context of dynamic crack-motion theory,\cite{RICE93COCH94} it was independently rediscovered in the context of the DPE by the author.\cite{PELL10} Albeit superficially dissipative-like and of the same form as the drag term (\ref{eq:dragstress}) it is non-phenomenological, and uniquely associated to $K$. It compensates for a contribution emanating from the latter in the equal-time limit $\Delta t\to 0$.\cite{PELL10,PELL11,PELL12} As it remains operative in steady motion, it should not be attributed the character of an acceleration/braking radiation term.

With the appropriate expression of $K$, Eq.\ (\ref{eq:pnstress}) applies to screw dislocations, or to edge dislocations of the ``glide'' type. For edge dislocation components of the ``climb" type\cite{WEER67a} the prefactor of the ``local" loss term differs from the one in (\ref{eq:pnstress}).\cite{PELL10,PERR95} For brevity, climb edge dislocations are not considered further hereafter.

As has been shown in Ref.\ \onlinecite{PELL12}, kernel $K$ is related to the steady-state quasimomentum function $p(v)$ of the dislocation where $v$ is a velocity,\cite{FRAN49,BELT68,HIRT98} by
\begin{equation}
\label{eq:kp}
K(x,t)=\frac{\theta(t)}{2 w_0}\lim_{\epsilon\to 0}\frac{e^{-\epsilon/t}}{t^2}p(x/t),
\end{equation}
where
\begin{equation}
\label{eq:w0}
w_0=\mu b^2/(4\pi)
\end{equation}
is a characteristic line energy density, and $\theta$ is the Heaviside function. The factor $e^{-\epsilon/t}$ regularizes the approach of $t=0$. Included here only for definiteness, it is needed when dealing with Volterra (zero-width core) dislocations but could be omitted in the present problem.\cite{PELL12} A slight abuse of language, repeated hereafter, has been committed in writing (\ref{eq:kp}) (see Note \onlinecite{NOTE4}). For a screw dislocation, $p(x/t)$ is a locally-integrable function; for an edge dislocation, one of its terms contains a ``finite part'' prescription.\cite{PELL12}

For uniform motion at velocity $v$ under constant stress, Eq.\ (\ref{eq:dyncore}) written in a co-moving Galilean frame reduces to\cite{PELL10}
\begin{eqnarray}
\label{eq:werteq}
-\frac{A(v)}{\pi}\int\frac{\dd x'}{x-x'}\frac{\partial\eta}{\partial x}(x')+B_\alpha(v)\frac{\partial\eta}{\partial x}(x')+\sigma_a=f'(\eta),\nonumber\\
\end{eqnarray}
where $B_\alpha(v)=B(v)+\alpha(\mu/2)v/\cS$, and where the integral is defined as a principal value at $x'=x$. This equation, first proposed by Weertman, who determined the functions $A(v)$ and $B(v)$,\cite{WEER69a,WEER69b} has more recently [with Eq.\ (\ref{eq:pullback})] been revisited by Rosakis under the name ``Model I".\cite{ROSA01}

The developments below rely on the availability of an explicit solution to Eq.\ (\ref{eq:werteq}). Such a solution is known when $\sigma_a$ is uniform and $|\siga|\leq\sigth$,\cite{ROSA01} which is why we restrict ourselves to such loadings. However, writing down the DPE with position-dependent $\sigma_a(x,t)$ is feasible, provided one adheres to definition (\ref{eq:ddens}) of the dislocation density rather than to the usual one. Under non-uniform $\sigma_a$, the background solution $\eta^{\rm e}$ becomes non uniform as well by Eq.\ (\ref{eq:etainfty}), so that defining $\rho$ as in Eq.\ (\ref{eq:ddens}) ensures that $\sigma_\eta$ is only due to the dislocation density, and possesses no spurious contribution from $\eta^{\rm e}$.

\section{Collective variables from d'Alembert's principle}
\label{sec:collvars}
\subsection{Method}
We know of no exact method of solution for the DPE. While multiple-dislocation solutions most certainly exist (just as Nabarro's dipole in statics\cite{NABA47}), the following application of the collective-variable method allows us to construct an approximate solution in the restricted subspace of single-dislocation solutions.

We start by reshuffling the degrees of freedom of $\eta$ by writing it as
\begin{eqnarray}
\label{eq:etaeta}
\eta(x,t)\equiv \eta_0(x,t)+\Delta\eta(x,t),
\end{eqnarray}
where $\eta_0$ is a single-dislocation ``mean-field'' ansatz, and $\Delta\eta$ is the residual. Likewise, we write the dislocation density as
\begin{eqnarray}
\rho(x,t)=\rho_0(x,t)+\Delta\rho(x,t)
\end{eqnarray}
where $\rho_0=(\partial\eta_0/\partial x)$ and $\Delta\rho=(\partial\Delta\eta/\partial x)$. According to the general principles of the CV approach, the ansatz must be of a form consistent with the steady-state limit of the field equation under study. We take therefore\cite{PELL12}
\begin{eqnarray}
\label{eq:etaansatz}
\eta_0(x,t)=\eta^{\rm e}(t)+\frac{b}{\pi}\left[\frac{\pi}{2}
-\mathop{\text{Arctan}}\frac{2\left(x-\xi(t)\right)}{a(t)}\right],
\end{eqnarray}
where $\eta^{\rm e}$ is given by Eq.\ (\ref{eq:etainfty}). The dislocation position along the $x$-axis, $\xi(t)$, and width, $a(t)$, stand as CVs for which governing equations are sought. In the steady state where $a$ is a constant, and in the co-moving frame where $x-\xi=x-vt$ is replaced by $x$, this ansatz solves Eq.\ (\ref{eq:werteq}) for non-supersonic velocities (i.e., $|v|<\cS$ for screws, and $|v|<\cL$ for edges),\cite{ROSA01} under conditions that connect $a$, $v$ and $\sigma_a$, to be retrieved in Sec.\ \ref{sec:steady}. It should be noted that (\ref{eq:etaansatz}) describes a dislocation with \emph{negative} density $\rho_0$. Calculations of a similar spirit have previously been carried out with a different formalism on a lattice dislocation model ---but with fixed width, the residual being further decomposed into phononlike degrees of freedom.\cite{NINO72}

By d'Alembert's principle,\cite{LANC86} a weak form of Eq.\ (\ref{eq:dyncore}) is obtained by requiring the virtual work to vanish for instantaneous variations $\delta\eta(x,t)$:
\begin{eqnarray}
\label{eq:varI}
\delta\mathcal{I}(t,[\eta])=\int\mathrm{d}x\, \mathcal{F}(x,t,[\eta])\,\delta\eta(x,t)\equiv 0.
\end{eqnarray}
The main governing equations for $\Delta\eta$ and the CVs simply follow from replacing in (\ref{eq:varI}) the field $\eta$ by its parametrization $\eta_0+\Delta\eta$, and by writing the variation $\delta\eta(x,t)$ in terms of the independent variations $\delta\Delta\eta(x,t)$, $\delta a(t)$, and $\delta\xi(t)$. Setting
\begin{equation}
\rho_1(x,t)=\frac{2}{a(t)}[x-\xi(t)]\rho_0(x,t),
\end{equation}
one obtains
\begin{align}
\label{eq:vareta}
\delta\eta&=\delta\Delta\eta-\rho_0\,\delta\xi-\rho_1\,\frac{\delta a}{a}.
\end{align}
Employing this expression in $\delta\mathcal{I}$ and zeroing each variation yields the following coupled equations of motion, expressed for convenience using the bracket notation $\braket{f_1}{f_2}=\int \dd x\, f_1(x)f_2(x)$, where $f_{1,2}$ are arbitrary functions:
\begin{subequations}
\label{eq:equsmot}
\begin{align}
\label{eq:equsmotde}
&\mathcal{F}(x,t,[\eta_0+\Delta\eta])=0,\\
\label{eq:equsmotaxi}
&\braket{\rho_i}{\mathcal{F}([\eta_0+\Delta\eta])}=0,\quad i=0,1.
\end{align}
\end{subequations}
Equation (\ref{eq:equsmotde}) determines $\Delta\eta$ given the CVs, whereas the set (\ref{eq:equsmotaxi}) provides governing equations for the CVs given the residual $\Delta\eta$. Equation (\ref{eq:equsmotde}) is nothing but the DPE (\ref{eq:dyncore}), in which $\eta$ has been substituted by $\eta_0+\Delta\eta$. On the other hand, Eqs.\ (\ref{eq:equsmotaxi}) are projections of (\ref{eq:equsmotde}) onto $\rho_0=-\partial_\xi\eta_0$ and $\rho_1=-\partial_a\eta_0$. Quite generally in the projector approach, the basis functions appear as derivatives of the ansatz in the collective coordinates.\cite{BOES88} The above derivation by d'Alembert's principle makes this obvious. Definition (\ref{eq:etaansatz}) of $\eta_0$ moreover implies the orthogonality property $\braket{\rho_0}{\rho_1}=0$.

To ensure equivalence between Eqs.\ (\ref{eq:equsmot}) and the DPE, the overall number of degrees of freedom must be preserved. Therefore, constraints must be imposed to relate $\Delta\eta$ to $a(t)$ and $\xi(t)$. Equations (\ref{eq:equsmotde}) and (\ref{eq:equsmotaxi}) are not yet usable, as they are unconstrained. The constraints are deduced\cite{BOES88,BOES90} from minimizing over $\xi$ and $a$ the quadratic norm of the residual\cite{TCHO93}
\begin{equation}
N=\int\dd x\,\Delta\eta(x,t)^2=\int\dd x\,[\eta(x,t)-\eta_0(x,t;a,\xi)]^2,
\end{equation}
where we have explicitly indicated the dependence of the ansatz in the CVs. Differentiating $N$ with respect to the CVs leads to the following constraints, to be obeyed at all times:
\begin{equation}
\label{eq:constr}
C_i(t)\equiv\braket{\Delta\eta}{\rho_i}\equiv 0,\qquad i=0,1.
\end{equation}
We shall assume that the initial state is either rest, or more generally a steady state at constant velocity $v$ [for which $\xi=vt$ and $a$ can be determined from the functions $A(v)$ and $B_\alpha(v)$; see (\ref{eq:av})]. Then $\rho_0(x,t)$ is the \emph{exact} solution at $t=0$. Consequently, $\Delta\eta=0$ and Eqs.\ (\ref{eq:constr}) are trivially satisfied at $t=0$. To enforce them at later times, it suffices to require their time derivative (denoted with a dot) to vanish, namely, $\dot{C}_{0,1}\equiv 0$.\cite{BOES90} This provides the following relationships between the first time derivative of $\Delta\eta(x,t)$ and that of the CVs:
\begin{equation}
\label{eq:dconstr}
\braket{\rho_i}{\partial_t\Delta\eta}+\braket{\partial_a\rho_i}{\Delta\eta}\dot{a}
+\braket{\partial_\xi\rho_i}{\Delta\eta}\dot{\xi}=0,\quad i=0,1.
\end{equation}
In short, the procedure to write down the constrained form of Eqs.\ (\ref{eq:equsmotde}) and (\ref{eq:equsmotaxi}) is as follows (see Appendix \ref{sec:evoleq} for details). Calling ``passive" in some equation a variable with no time derivative thereof involved, the principle is to make $\Delta\eta$ ``passive" in the equations for the CVs, and to make the CVs ``passive" in the equation for $\Delta\eta$. To achieve this, one expresses in (\ref{eq:equsmotde}) in terms of $\partial_t\Delta\eta$ the time derivatives of the CVs arising from $\partial_t\eta_0$; conversely, in Eqs.\ (\ref{eq:equsmotaxi}), one expresses $\braket{\rho_i}{\partial_t\Delta\eta}$ in terms of the time derivatives of the CVs. These substitutions are carried out by means of (\ref{eq:dconstr}). This procedure is adequate to equations involving only first-order time derivatives such as the DPE. Were higher-order derivatives present, further substitutions would be needed, differentiating Eq.\ (\ref{eq:dconstr}) with respect to $t$ to obtain the necessary relationships.\cite{BOES88}

In the rest of the paper, we consider Eqs.\ (\ref{eq:equsmotaxi}) only to leading order in $\Delta\eta$, leaving to further work the study of the residual and of its influence on the CVs. Then, constraints (\ref{eq:dconstr}) and Equ.\ (\ref{eq:equsmotde}) are irrelevant (see Appendix \ref{sec:evoleq}) and the equations reduce to
\begin{subequations}
\label{eq:xiao}
\begin{eqnarray}
\label{eq:xio}
&&\braket{\rho_0}{\mathcal{F}[\eta_0]}=\int\mathrm{d}x\,\rho_0\,\mathcal{F}[\eta_0]=0,\\
\label{eq:ao}
&&\braket{\rho_1}{\mathcal{F}[\eta_0]}=\int\mathrm{d}x\,\rho_1\,\mathcal{F}[\eta_0]=0.
\end{eqnarray}
\end{subequations}
Equation (\ref{eq:xio}) has been studied in Ref.\ \onlinecite{PELL12} where, following Eshelby,\cite{ESHE53} it was postulated (rather than derived as above). By contrast, Equ.\ (\ref{eq:ao}) has not previously been considered for dynamic dislocation motion. Since $\mathcal{F}$ is a stress and $\rho_0$ is a Burgers ``charge" density, Eq.\ (\ref{eq:xio}) represents a dynamic force balance equation. Likewise, Eq.\ (\ref{eq:ao})\----a virial-type equation\----\-expresses a dynamic energy balance.

\subsection{Comments}
Some general comments are in order, since Boesch \textit{et al.}'s systematic CV theory was developed in the framework of Lagrange-Hamilton dynamics (LHD).\cite{BOES88,BOES90} As such, the original approach is well-suited to propagating kink models such as the sine-Gordon one, or Frenkel-Kontorova's,\cite{PEYR04,BRAU04} which are one-dimensional from the outset and admit a Hamiltonian.\cite{BRAU04} By contrast (see Introduction) the DPE results from a one-dimensional reduction of a two-dimensional elastodynamic problem, which makes it his\-to\-ry-de\-pen\-dent and dissipative, and elude standard Hamiltonian dynamics. This prompted us to use d'Alembert's priciple instead. To make the connection with the original CV method, we note first that there is of course complete equivalence between d'Alembert's principle and standard LHD in non-dissipative cases where the latter approach can be used: starting from kinetic and potential energies, LHD provides governing force-balance equations that can be cast in the variational form of d'Alembert's principle as in Eq.\ (\ref{eq:varI}); conversely, given non-dissipative and non-history-dependent field equations, Hamilton's variational principle can be logically deduced from d'Alembert's principle.\cite{LANC86} However, d'Alembert's principle is more fundamental in the sense that it is operative without any restriction.\cite{LANC86} Second, Boesch \textit{et al.} showed within LHD that using the constraints in the form of total time derivatives as in Eq.\ (\ref{eq:dconstr}) alleviates the need for Lagrange multipliers, which we need not use either. Indeed, adding such constraints by means of Lagrange multipliers to some hypothetical Lagrangian would leave the Euler-Lagrange equations of motion unchanged, the constraints acting as an ignorable null Lagrangian.\cite{BOES90} Thus, the same governing equations for the CVs and the residual as with d'Alembert's principle would be obtained, the constraints being put into action in both cases as above, namely, in a second step by substitutions in the governing equations. This shows that the present approach is fully consistent with that by Boesch \textit{et al.}, while being usable with the DPE for which no Lagrangian is available, mainly due to the ``local" term in Eq.\ (\ref{eq:pnstress}).
\section{Equations of motion}
\label{sec:eom}
\subsection{Governing equation for $\xi(t)$}
\label{sec:eqpos}
We first briefly recall the explicit form of Equ.\ (\ref{eq:xio}) for $\xi$, already obtained in Ref.\ \onlinecite{PELL12}, which we cast hereafter in a slightly different form. For notational consistency with the latter work we drop from now on the subscript ${}_0$ in  $\eta_0(x,t)$ and $\rho_0(x,t)$ unless otherwise stated and denote these quantities by  $\eta(x,t)$ and $\rho(x,t)$, keeping in mind that they refer to ansatz (\ref{eq:etaansatz}). Compatibility with the latter requires us to restrict ourselves to homogeneous stress conditions $\sigma_a(x,t)\equiv \sigma_a(t)$. Indeed, $\eta^{\rm e}(t)$ can be independent of $x$ only if $\sigma_a$ is. Introduce the complex position-width coordinate ($\ii=\sqrt{-1}$)
\begin{equation}
\zeta(t)=\xi(t)+\ii\frac{a(t)}{2},
\end{equation}
and the mean complex ``velocity" between instants $\tau$ and $t$
\begin{equation}
\label{eq:vbar}
\overline{v}(t,\tau)=\frac{\zeta(t)-\zeta^*(\tau)}{t-\tau},
\end{equation}
where the star denotes the complex conjugate. In Ref.\ \onlinecite{PELL12}, Eq.\ (\ref{eq:xio}) was reduced to
\begin{eqnarray}
\label{eq:eomxip}
-2\Re\int_{-\infty}^t\frac{\dd\tau}{\Delta t^2}p(\overline v)+\kappa\frac{2w_0}{\cS a(t)}\dot{\xi}(t)-b\sigma_a(t)=0,
\end{eqnarray}
where $\Delta t=t-\tau$, and $\kappa=1+\alpha$. The quantity $\cS$ is the shear wave velocity, $\overline{v}$ stands for $\overline{v}(t,\tau)$, and $p(v)$ is the quasimomentum function relevant to screw or edge dislocations introduced in Eq.\ (\ref{eq:kp}). The equal-time limit of $p(\overline{v})$ is purely imaginary if $a\not=0$, of value
\begin{equation}
\label{eq:piinfty}
p(\overline{v}(t,t))=\lim_{\tau\to t^-}p(\overline{v}(t,\tau))=p(+\ii\infty)=\ii\frac{w_0}{\cS}.
\end{equation}
More precisely,
\begin{equation}
\label{eq:pexpand}
p(\overline{v}(t,\tau))=p(\overline{v}(t,t))+O(\Delta t^2).
\end{equation}
To underline the connection with Eq.\ (\ref{eq:eomadp}) below, it is appropriate to introduce the quasimomentum variation
\begin{equation}
\Delta p(t,\tau)= p(\overline{v}(t,t))- p(\overline{v}(t,\tau)),
\end{equation}
and write (\ref{eq:eomxip}) as
\begin{eqnarray}
\label{eq:eomxidp}
2\Re\int_{-\infty}^t\dd\tau\frac{\Delta p}{{\Delta t^2}}+\kappa\frac{w_0}{\cS}\frac{2}{a}\dot{\xi}-b\sigma_a=0,
\end{eqnarray}
where $\dot\xi$, $a$ and $\sigma_a$ are evaluated at instant $t$. Because of the $\Re$ operator, this modification is only a ``cosmetic" one. Introducing next the mass function
\begin{equation}
m(v)=\frac{\dd p}{\dd v}(v),
\end{equation}
and integrating by parts the first term of (\ref{eq:eomxidp}), the boundary contribution at $\tau=-\infty$ vanishes trivially, while that at $\tau=t$ vanishes owing to (\ref{eq:pexpand}). One thus arrives at the governing equation for $\xi$ in ``mass form'',
\begin{eqnarray}
\label{eq:eomxim}
2\Re\int_{-\infty}^t\frac{\dd\tau}{\Delta t}m(\overline{v})\frac{\dd\overline{v}}{\dd\tau}+\kappa\frac{w_0}{\cS}\frac{2}{a}\dot{\xi}-b\sigma_a=0.
\end{eqnarray}
Its most important component, the \emph{self-force}, is the sum of the first two terms with $\kappa=1$. It has been studied in Ref.\ \onlinecite{PELL12}.

\subsection{Governing equation for $a(t)$}
\label{sec:eqwidth}
Turning to (\ref{eq:ao}), we evaluate in succession each contribution to $(2/a)\int\mathrm{d}x\, (x-\xi)\rho\,\mathcal{F}$, with $\mathcal{F}$ read from Eqs.\ (\ref{eq:fdef})--(\ref{eq:dragstress}). Consider first
\begin{align}
\frac{2}{a(t)}\int_{-\infty}^{+\infty}&\mathrm{d}x\,[x-\xi(t)]\rho(x,t)[\sigma_a(t)-f'(\eta(x,t))]\nonumber\\
\label{eq:intsig}
&=b\sigma_{\text{th}}\sqrt{1-\sigma_a(t)^2/\sigma_{\text{th}}^2},
\end{align}
where expression (\ref{eq:etainfty}) has been used. Next,
\begin{eqnarray}
\frac{2}{a(t)}\int\mathrm{d}x\,[x-\xi(t)]\rho(x,t)\frac{\partial\widetilde{\eta}}{\partial t}(x,t)=-\frac{b^2}{2\pi}\frac{\dot{a}(t)}{a(t)}.
\end{eqnarray}
Finally, one finds that, for both screw and edge,
\begin{eqnarray}
&&-\frac{2}{a(t)}\frac{\mu}{\pi}\int\mathrm{d}x\,\mathrm{d}x'\,[x-\xi(t)]\rho(x,t)K(x,t|x',\tau)\rho(x',\tau)\nonumber\\
\label{eq:imdeltap}
&&{}=-\frac{2}{\Delta t^2}\Im \Delta p(t,\tau).
\end{eqnarray}
The calculation leading to (\ref{eq:imdeltap}) closely follows the Fourier-transform approach detailed in Ref.\ \onlinecite{PELL12} and uses the integrals provided in that reference. Gathering terms yields the governing equation for $a(t)$,
\begin{eqnarray}
\label{eq:eomadp}
&&{}-2\Im\int_{-\infty}^t \dd\tau \frac{\Delta p}{\Delta t^2}+\kappa\frac{w_0}{\cS}\frac{\dot{a}}{a}+b\sigma_{\rm th}\sqrt{1-\frac{\sigma_a^2}{\sigma_{\rm th}^2}}=0,
\end{eqnarray}
or, in ``mass form'', after integrating by parts, and changing the sign
\begin{eqnarray}
\label{eq:eomam}
2\Im\int_{-\infty}^t \frac{\dd\tau}{\Delta t} m(\overline{v})\frac{\dd\overline{v}}{\dd\tau}-\kappa\frac{w_0}{\cS}\frac{\dot{a}}{a}-b\sigma_{\rm th}\sqrt{1-\frac{\sigma_a^2}{\sigma_{\rm th}^2}}=0.\nonumber\\
\end{eqnarray}

\subsection{Combined governing equation for $\zeta(t)$}
\label{sec:synth}
Equations (\ref{eq:eomxim}) and (\ref{eq:eomam}) are seen to constitute the real and imaginary parts of one single complex EoM for $\zeta(t)$, namely,
\begin{subequations}
\label{eq:eomzetam12}
\begin{eqnarray}
2\int_{-\infty}^t \frac{\dd\tau}{\Delta t} m(\overline{v})\frac{\dd\overline{v}}{\dd\tau}+\kappa\frac{w_0}{\cS}\frac{\dot{\zeta}^*}{\Im\zeta}=-\ii\,b\sigma_{\rm th}g\left(\frac{\sigma_a}{\sigma_{\rm th}}\right),\nonumber\\
\label{eq:eomzetam}
&&
\end{eqnarray}
where
\begin{equation}
\label{eq:gz}
g(x)=-\sqrt{1-x^2}+\ii x\qquad (|x|\leq 1).
\end{equation}
\end{subequations}
 The \emph{generalized (complex) self-force} associated to $\zeta(t)$ is
\begin{equation}
\label{eq:sfzeta}
F_\zeta(t)=2\int_{-\infty}^t \frac{\dd\tau}{\Delta t} m(\overline{v})\frac{\dd\overline{v}}{\dd\tau}+\frac{w_0}{\cS}\frac{\dot{\zeta}^*}{\Im\zeta}.
\end{equation}
The dislocation position and half-width are deduced from $\zeta(t)$ as $\xi(t)=\Re\zeta(t)$ and $a(t)/2=\Im\zeta(t)$. Equation (\ref{eq:eomzetam}) is equivalent to
\begin{equation}
\label{eq:eomzetamfund}
\int\mathrm{d}x\, (x-\zeta^*)\rho_0\,\mathcal{F}[\eta_0]=0,
\end{equation}
which follows from combining (\ref{eq:xio}) and (\ref{eq:ao}).

EoM (\ref{eq:eomzetam}) is a retarded integro-differential functional equ\-a\-tion of a type unprecedented for dislocations. The occurrence of complex numbers allows it to deal with faster-than-wave motion without any modification. This technical simplification finds its origin in the simple Lorentzian form of the ansatz density $\rho_0(x,t)$, which has one pair of conjugate poles $x=\xi(t)\pm\ii a(t)/2$. However, the physical significance of the complex-valued nature of the EoM is not obvious, although a connection between imaginary parts and dissipation exists (next Section). The presence of the ``mean velocity'' $\overline{v}$ instead of the instantaneous velocity\cite{BELT68} in the mass kernel is more transparent, as it indicates that retarded self-interactions are mediated by elastic waves between past emission times $\tau$ at position $\xi(\tau)$  and current time $t$ at position $\xi(t)$.\cite{ESHE53,PILL07,PELL12}

\subsection{Steady motion}
\label{sec:steady}
We examine next the steady-state form of (\ref{eq:eomzetam12}). It is obtained by assuming that $\zeta(t)=v t+\ii (a/2)$, where the dislocation velocity $v$ and the dislocation width $a$ are constant. Then,
\begin{subequations}
\begin{eqnarray}
\label{eq:vbarsteady}
\overline{v}(t,\tau)&=&v+\ii\frac{a}{\Delta t},\\
\label{eq:dvbarsteady}
\frac{\dd\overline{v}}{\dd\tau}(t,\tau)&=&\ii\frac{a}{\Delta t^2}.
\end{eqnarray}
\end{subequations}
The integral in (\ref{eq:eomzetam}) can be carried out exactly by changing the integration variable into $u=\overline{v}(t,\tau)$, and by remarking that
\begin{equation}
m(v)=\frac{\dd p(v)}{\dd v}=\frac{\dd^2 L(v)}{\dd v^2},
\end{equation}
where $L(v)$ is the \emph{steady-state} Lagrangian built from the elastic field of the dislocation.\cite{HIRT98,NOTE4} While obviously a related object, this quantity is \emph{not} the Lagrangian function of the model in the sense of Hamiltonian dynamics. In terms of $u$,
\begin{equation}
\frac{1}{\Delta t}=\frac{1}{\ii a}(u-v),
\end{equation}
so that
\begin{eqnarray}
&&2\int_{-\infty}^t \frac{\dd\tau}{\Delta t} m(\overline{v})\frac{\dd\overline{v}}{\dd\tau}=\frac{2}{\ii a}\int_{v+\ii0^+}^{+\ii\infty}\dd u\,(u-v)\frac{\dd^2 L}{\dd u^2}\nonumber\\
&=&\frac{2}{\ii a}\left\{[(u-v)p(u)]^{\ii\infty}_{v+\ii 0^+}-\int_{v+\ii0^+}^{+\ii\infty}\dd u\,\frac{\dd L}{\dd u}\right\}\\
&=&\frac{2}{\ii a}\left\{\lim_{u\to+\ii\infty}[(u-v)p(u)-L(u)]+L(v+\ii0^+)\right\}.\nonumber
\end{eqnarray}
The function $W(v)=v\,p(v)-L(v)$ is the steady-state line energy density,\cite{HIRT98,NOTE4} and it has been shown in Ref.\ \onlinecite{PELL12} that $W(+\ii\infty)=0$ for both screws and edges. Invoking moreover (\ref{eq:piinfty}) leads to
\begin{equation}
\label{eq:intsteady}
2\int_{-\infty}^t \frac{\dd\tau}{\Delta t} m(\overline{v})\frac{\dd\overline{v}}{\dd\tau}=
\frac{2}{\ii a}L(v+\ii0^+)-\frac{w_0}{\cS}\frac{2}{a}v.
\end{equation}
The steady-state expression of the self-force (\ref{eq:sfzeta}) follows as
\begin{equation}
\label{eq:selfss}
F_\zeta(v)=\frac{2}{\ii a(v)}L(v+\ii 0^+).
\end{equation}
The phenomenological drag term can be included in an \emph{augmented Lagrangian} defined as
\begin{equation}
\label{eq:lalp}
L_\alpha(v)=L(v+\ii 0^+)+\ii \alpha w_0\frac{v}{\cS}
\end{equation}
The real-valued functions $A(v)$ and $B_\alpha(v)$ in Eq.\ (\ref{eq:werteq}) are related to the real and imaginary parts of $L_\alpha(v)$ by the identity\cite{PELL12}
\begin{align}
L_\alpha(v)=2w_0\left[-A(v)+\ii B_\alpha(v)\right].
\end{align}
Non-zero values of $\Im L(v+\ii 0^+)$ arise for transonic ($\cS<|v|<\cL$, for  edges only) or supersonic ($|v|>\cS$ for screws; $|v|>\cL$ for edges) velocities in connection with dissipation in Mach fronts. Overall, steady-state dissipation processes are described by $\Im L_\alpha$.

\begin{figure}[!ht]
\includegraphics[width=7.00cm]{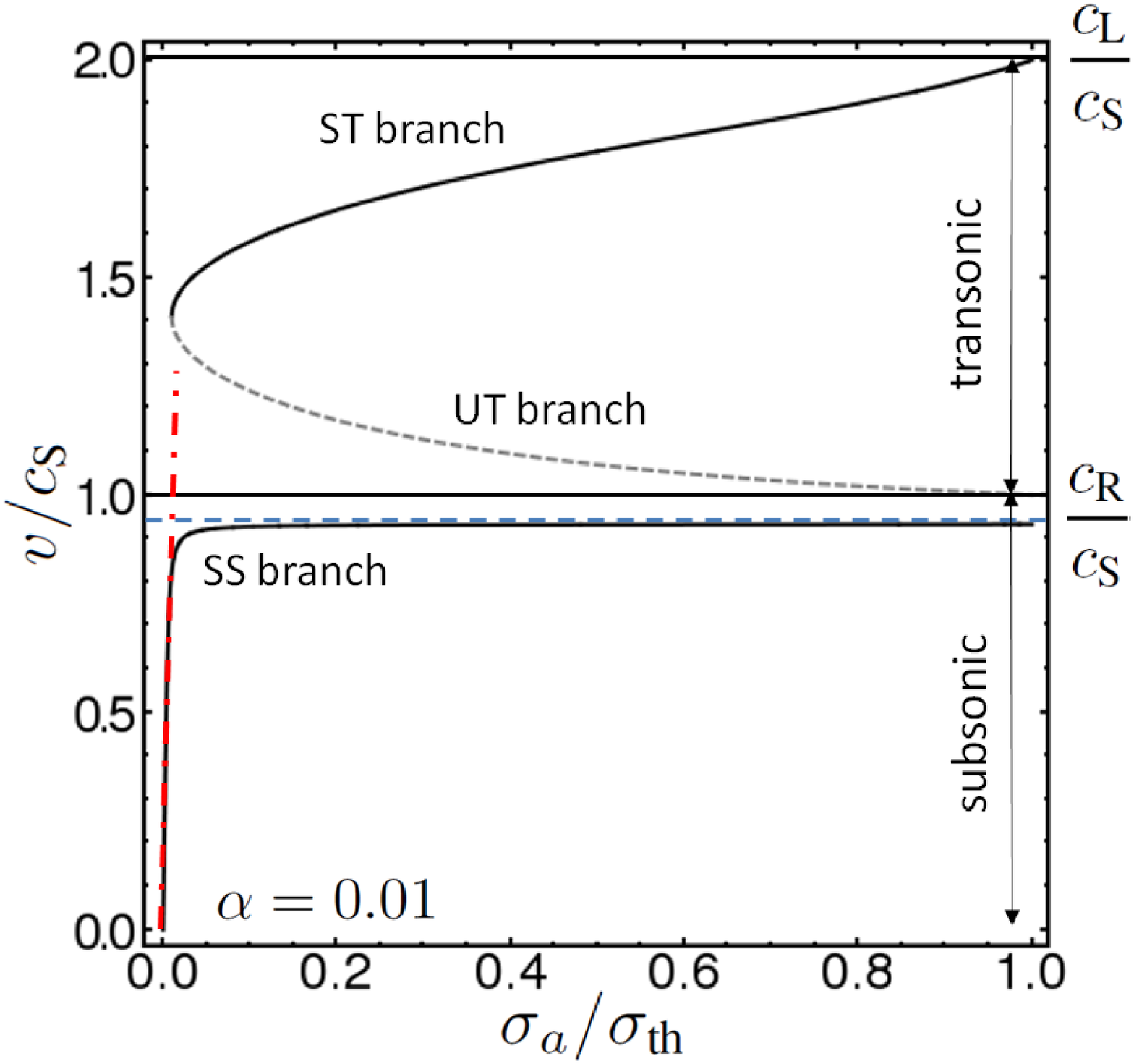}
\caption{\label{fig:fig1}(Color online) Edge dislocation. Velocity/stress branches of the steady-state relationship (\ref{eq:sv}) for $\alpha=0.1$ and $\cL=2\cS$. Solid black: stable branches; dashed grey: unstable branch. Dot-dashed (red): linear regime (\ref{eq:linreg}). Horizontal lines: limiting velocities $\cL$, $\cS$, and Rayleigh velocity $\cR$. The interval $(\cR,\cS)$ constitutes a velocity gap.}
\end{figure}

Since $\kappa=1+\alpha$ the steady-state form of the left-hand side of (\ref{eq:eomzetam}), namely, the sum of the generalized self-force (\ref{eq:selfss}) and the drag force, reads
\begin{equation}
\label{eq:selfss2}
F_\alpha(v)\equiv\frac{2L_\alpha(v)}{\ii a(v)},
\end{equation}
whereby EoM (\ref{eq:eomzetam}) reduces to
\begin{equation}
\label{eq:steadyf}
F_\alpha(v)=-\ii\,b\,\sigma_{\rm th}g\left(\frac{\sigma_a}{\sigma_{\rm th}}\right).
\end{equation}
Combining (\ref{eq:w0}) and (\ref{eq:sigth}) yields the identity $b\sigma_{\rm th}/w_0=2/d$, which with (\ref{eq:selfss2}) brings Eq.\ (\ref{eq:steadyf}) to
\begin{equation}
\label{eq:steady}
\frac{d}{w_0}L_\alpha(v)=a\,g\left(\frac{\sigma_a}{\sigma_{\rm th}}\right).
\end{equation}
Remarking that $|g(x)|=1$ by (\ref{eq:gz}), the width $a(v)$ follows from taking the modulus of (\ref{eq:steady}):
\begin{subequations}
\label{eq:steadystate}
\begin{equation}
\label{eq:av}
a(v)=\frac{d}{w_0}|L_\alpha(v)|.
\end{equation}
Then, (\ref{eq:steady}) reduces to a condition of equality between the complex arguments of both sides. It provides the stress-velocity relationship
\begin{equation}
\label{eq:sv}
\sigma_a=\sigth \sin\mathop{\rm Arg}L_\alpha(v)\quad \bigl(\cos\mathop{\rm Arg}L_\alpha(v)\leq 0\bigr).
\end{equation}
\end{subequations}
Equations (\ref{eq:av}) and (\ref{eq:sv}) are a reformulation, in complex La\-gran\-gian form, of the steady-state kinetic relations of Rosakis's Model I.\cite{ROSA01} The above shows that Eq.\ (\ref{eq:eomzetam}) stands as a leading-order approximation to the fully dynamic extension of this model. In Ref.\ \onlinecite{PELL12}, for lack of the governing equation (\ref{eq:eomam}), an instantaneous relationship $a(t)\equiv a\bigl(\dot{\xi}(t)\bigr)$ with $a(v)$ given by (\ref{eq:av}) was assumed in order to complement (\ref{eq:eomxim})\----\-using different notations. Employing (\ref{eq:eomzetam}) avoids this approximation, and \emph{produces} Eqs.\ (\ref{eq:av}) and (\ref{eq:sv}) as particular steady-state consequences. Figure \ref{fig:fig1} represents, for an edge dislocation, the velocity/stress relationship deduced from inverting (\ref{eq:sv}), where we introduce the denominations `stable subsonic' (SS), 'stable transonic' (ST), and 'unstable transonic' (US) branches, to be used hereafter. The unstable branch is characterized by $\dd v/\dd\siga <0$.\cite{ROSA01}

The content of Equ.\ (\ref{eq:steady}) is as follows. The dislocation core is a region of high atomic disregistry akin to a stacking fault spread over the width $a$. In expanded form, and setting for brevity $s=\sigma_a/\sigth$ Equ.\ (\ref{eq:steady}) reads
\begin{equation}
\label{eq:content}
2w_0[-A(v)+\ii B_\alpha(v)]=w_0 \frac{a}{d}\left[-\sqrt{1-s^2}+\ii s\right],
\end{equation}
in which $(a/d)w_0\sqrt{1-s^2}$ represents the misfit energy cost of the core, given the application of $\sigma_a$. Moreover, by definition of the Lagrangian function, $2w_0 A(v)$ is the excess elastic strain energy, over the displacement kinetic energy, at velocity $v$. Denoting by $\cR$ the Rayleigh velocity [root of $A(v)$ for an edge dislocation], the restriction in (\ref{eq:sv}) constrains this energy difference to be non-negative, and forbids edge dislocations to move steadily in the velocity gap $(\cR,\cS)$ (see Fig.\ \ref{fig:fig1}), the only interval where it is negative. The excess energy determines the misfit energy cost. Moreover, the imaginary part of Equ.\ (\ref{eq:content}) provides the mobility law as an equilibrium equation between the Peach-Koehler force $b\sigma_a$ and the overall drag force, in the form
\begin{equation}
\label{eq:mobi}
b\sigma_a=2 (d b/a) B_\alpha(v)=(d b/a)\Im L_\alpha(v)/w_0.
\end{equation}

The prescription $+\ii 0^+$ in Eq.\ (\ref{eq:lalp}) has the following meaning. For the dislocation to move in the direction of the applied stress, $v$ must be of the sign of $\siga$. Thus, by Eqs.\ (\ref{eq:lalp}) and (\ref{eq:mobi}), $\Im L(v)$ must be an odd function. The infinitesimal quantity $+\ii 0^+$ handles the branch cut of the square-root ``relativistic" factors in $L(v)$ for transonic or supersonic motion to fulfill this condition. Its introduction finds its origin in the core width being nonzero, which can therefore be considered a necessary requirement for transitions to faster-than-wave states.

For small applied stress and velocity (Fig.\ \ref{fig:fig1}), `relativistic' effects radiative drag is negligible and the mobility law (\ref{eq:mobi}) reduces to the well-known linear law
\begin{equation}
\label{eq:linreg}
\widetilde{B} v=b\sigma_a,
\end{equation}
where $\widetilde{B}$ is the usual drag coefficient [not to be confused with $B(v)$] expressed in units of Pa.s. In terms of $\alpha$, $\widetilde{B}$ reads
\begin{equation}
\label{eq:balpha}
\widetilde{B}\equiv \frac{2 w_0}{a(0)\cS}\alpha=\frac{\mu b^2}{2\pi a(0)\cS}\alpha,
\end{equation}
where $a(0)$ is the core width at rest. For a screw (resp., edge) dislocation Equ.\ (\ref{eq:av}) gives $a(0)=d$ [resp., $d/(1-\nu)$, where $\nu$ is the Poisson ratio]. Both $b$ and the interplane distance $d$ depend on the slip system.\cite{HIRT82} Use of (\ref{eq:balpha}) with values of $\widetilde{B}$ from atomistic simulations\cite{OLMS05} at 300 K on Al and Ni provides values of $\alpha$ of typical order of magnitude $10^{-2}$.

\section{Local analysis of the self-force}
\label{sec:local}
Further insight into the generalized self-force (\ref{eq:sfzeta}) is obtained by carrying out an expansion of the integrand near the current time $\tau=t$. To this aim, we assume that accelerated or decelerated motion begins at time $\tau=0$ and position $\xi(0)=0$, following initial steady-state motion (possibly rest) from $\tau=-\infty$ to $\tau=0$ with constant velocity $\dot{\zeta}=\vi$ and core width $\ai$. The force is split as
\begin{equation}
F_\zeta(t)=F_\zeta^<(t)+F_\zeta^>(t),
\end{equation}
where, for $t>0$,
\begin{subequations}
\begin{eqnarray}
\label{eq:finf}
F_\zeta^<(t)&=&2\int_{-\infty}^0 \frac{\dd\tau}{\Delta t} m(\overline{v})\frac{\dd\overline{v}}{\dd\tau},\\
\label{eq:sup}
F_\zeta^>(t)&=&2\int_0^t \frac{\dd\tau}{\Delta t} m(\overline{v})\frac{\dd\overline{v}}{\dd\tau}+\frac{w_0}{\cS}\frac{\dot{\zeta}^*}{\Im\zeta}.
\end{eqnarray}
\end{subequations}
The term $F_\zeta^<(t)$ is computed in closed form by the method of Sec.\ \ref{sec:steady}. One obtains\cite{PELL12}
\begin{subequations}
\label{eq:finfexpr0}
\begin{equation}
\label{eq:finfexpr}
F_\zeta^<(t)=\frac{2}{t}\left[p\left(v^<(t)\right)-\frac{L(v^<(t))-L(\vi+\ii 0^+)}{v^<(t)-\vi}\right],
\end{equation}
where
\begin{equation}
\label{eq:vinf}
v^<(t)\equiv \frac{1}{t}\left[\zeta(t)+\ii\frac{\ai}{2}\right]=\frac{\xi(t)}{t}+\ii\frac{[a(t)+\ai]}{2t}.
\end{equation}
\end{subequations}
The series expansion of $F_\zeta^>(t)$ proceeds from
\begin{equation}
\zeta(\tau)=\zeta(t)-\dot{\zeta}(t)\Delta t+\frac{1}{2}\ddot{\zeta}(t)\Delta t^2
-\frac{1}{6}\dddot{\zeta}(t)\Delta t^3+O\left(\Delta t^4\right).
\end{equation}
Introducing
\begin{equation}
u(t,\tau)\equiv\dot{\zeta}^*(t)+\ii\frac{a(t)}{\Delta t},
\end{equation}
one finds
\begin{subequations}
\begin{eqnarray}
\vbar(t,\tau)
&=&[\zeta(t)-\zeta^*(\tau)]/\Delta t\\
\label{equ:exvbar}
&=&u(t,\tau)-\frac{1}{2}\ddot{\zeta}^*(t)\Delta t+\frac{1}{6}\dddot{\zeta}(t)^*\Delta t^2+O\left(\Delta t^3\right),\nonumber
\end{eqnarray}
whence
\begin{equation}
\frac{\dd\vbar}{\dd\tau}(t,\tau)
=\frac{\ii a(t)}{\Delta t^2}+\frac{1}{2}\ddot{\zeta}^*(t)\Delta t-\frac{1}{3}\dddot{\zeta}^*(t)\Delta t+O\left(\Delta t^2\right).
\end{equation}
\end{subequations}
Expanding next $m(\vbar)$ around $\vbar=u$ yields, with $\zeta=\zeta(t)$ and time derivatives of $\zeta^*$ evaluated at $t$
\begin{eqnarray}
2m(\vbar)
&=&2 m(u)-m'(u)\ddot{\zeta}^*\Delta t\\
&&{}+\left[\frac{1}{3}m'(u)\dddot{\zeta}^*
+\frac{1}{4}m''(u)\ddot{\zeta}^{*2}\right]\Delta t^2+O\left(\Delta t^2\right).\nonumber
\end{eqnarray}
Using the above expansions provides, with $a=a(t)$,
\begin{eqnarray}
\label{eq:expan}
\frac{2}{\Delta t}m(\vbar)\frac{\dd\vbar}{\dd\tau}
&=&
\frac{1}{\Delta t^3}\Bigl\{
2\ii a\,m(u)\ell^0-\ii a\, m'(u)\ddot{\zeta}^*\Delta t\,\ell^1\nonumber\\
&+&\Bigl[m(u)\ddot{\zeta}^*+\ii\frac{a}{3}m'(u)\dddot{\zeta}^*\nonumber\\
&+&\ii\frac{a}{4}m''(u)\ddot{\zeta}^{* 2}\Bigr]\Delta t^2\,\ell^2
\Bigr\}+O\left(\Delta t^0\ell^3\right),
\end{eqnarray}
where a bookkeeping variable $\ell$, to be taken as $\ell=1$ in the final result, has been introduced in the numerator to keep track of the expansion order of the terms.

The expansion of $F_\zeta^>$ follows from integrating each term of (\ref{eq:expan}) over $\tau$ from $\tau=0$ to $\tau=t^-$. As $u(t,\tau)$ is of the form (\ref{eq:vbarsteady}) integrations can be carried out as in the previous Section by using $u$ as an integration variable on a path in the upper complex half-plane, going from $u=v^>(t)$ to $u=+\ii\infty$, where
\begin{subequations}
\label{eq:fsupexpansion}
\begin{equation}
\label{eq:vsup}
v^>(t)\equiv\dot{\zeta}^*(t)+\ii\frac{a(t)}{t}.
\end{equation}
The result reads, with $v^>=v^>(t)$,
\begin{eqnarray}
F_\zeta^>(t)&=&\frac{2}{\ii a}\left[L\left(v^>\right)-\ii\frac{a}{t}p\left(v^>\right)\right]\ell^0+m\left(v^>\right)\ddot{\zeta}^*\ell^1\nonumber\\
&&{}+\Bigl[M^{(0)}\left(\dot{\zeta}^*,\frac{a}{t}\right)\ddot{\zeta}^*
+\frac{\ii}{3}a\,M^{(1)}\left(\dot{\zeta}^*,\frac{a}{t}\right)\dddot{\zeta}^*\nonumber\\
\label{eq:fsupexp}
&&{}+\frac{\ii}{4}a\,M^{(2)}\left(\dot{\zeta}^*,\frac{a}{t}\right)\ddot{\zeta}^{*2}\Bigr]\ell^2
+O\left(\ell^3\right),
\end{eqnarray}
where the following functions, defined for $y>0$ and $v$ an arbitrary complex number, have been introduced:
\begin{eqnarray}
\label{eq:mvy}
M^{(0)}(v,y)&=&\int_y^{+\infty}\frac{\dd z}{z}m(v+\ii z),\\
M^{(k)}(v,y)&=&\frac{\partial^k M^{(0)}}{\partial v^k}(v,y),\qquad (k\geq 1).
\end{eqnarray}
\end{subequations}

To understand $F^>_\zeta$ and $F^<_\zeta$, consider an evolution between two steady states of different velocities. In the initial and final states, different stress fields surround the dislocation. In the course of motion, the ``old'' field is replaced by the ``new" one, which occurs via wave emission from the dislocation. Thus, the motion can be viewed as simultaneous steps of \emph{destruction} of the old field, and of \emph{creation} of the new one. Quite generally, the ``destruction" contribution is the part of the integral from $\tau=-\infty$ up to the instant where accelerated motion begins, taken as $\tau=0$ by convention. This is already apparent at the level of the field $\sigma_\eta$, e.g., for a Volterra screw dislocation [see Eq.\ (9) in Ref.\ \onlinecite{PELL12}]. In the CV framework, this destruction contribution is represented by $F_\zeta^<(t)$. It is purely kinematic, since Eqs.\ (\ref{eq:finfexpr0}) depend on $\xi(t)$ and $a(t)$ but not on their derivatives. However, (\ref{eq:finfexpr}) has an ``effective" inertial content since for $v\simeq \vi$,
\begin{equation}
\frac{2}{t}\left[p(v)-\frac{L(v)-L(\vi)}{v-\vi}\right]\simeq m(\vi)\frac{v-\vi}{t},
\end{equation}
where $(v-\vi)/t$ is akin to an acceleration.

The ``creation" part is described by $F_\zeta^>(t)$. In (\ref{eq:fsupexp}), the leading term involves velocities and $a(t)$ but not accelerations. In the steady-state limit $\dot{\xi}\to v$ at large times, it reduces to $2 L(v+\ii 0^+)/(\ii a)$, namely, expression (\ref{eq:selfss}). Dynamic transients are accounted for by the higher-order terms, which vanish in the latter limit. Among them is a Newtonian-like inertial term of the form $m^{\text{eff}}\ddot{\zeta}^*$, with effective mass
\begin{equation}
m^{\text{eff}}(\dot{\xi},a,\dot{a},t)=m(v^>)+M^{(0)}\left(\dot{\zeta}^*,\frac{a}{t}\right),
\end{equation}
obtained by combining terms of order $\ell^1$ and $\ell^2$. It is noted that integral (\ref{eq:mvy}) diverges logarithmically as $y\to 0$, so that $M^{(0)}(\zeta,a/t)$ displays at large times the well-known  $\ln t$ inertial behavior \cite{ESHE53} (see also, e.g., Ref.\ \onlinecite{PELL12}). The remaining terms of order $\ell^2$ involve the so-called jerk (third time derivative of motion), and the acceleration squared. The latter non-linear term is a consequence of ``relativistic-like'' effects. Notwithstanding imaginary parts, their sum represents the elastodynamic equivalent of the Lorentz-Abraham reaction-radiation force on charged classical particles in electrodynamics.\cite{JACK75}

The above suggests another decomposition of $F_\zeta$, namely,
\begin{equation}
F_\zeta(t)=F^{\text{dyn}}_\zeta(t)+F^{\text{adia}}_\zeta(t),
\end{equation}
where the ``dynamic" part $F^{\text{dyn}}_\zeta$ gathers the Newtonian-like and higher-order terms in $F^>_\zeta$, and where the ``adiabatic" part,
\begin{eqnarray}
F^{\text{adia}}_\zeta&=&\frac{2}{t}\left[
p\left(v^<\right)-p\left(v^>\right)-\frac{L(v^<)-L(\vi+\ii 0^+)}{v^<-\vi}\right]\nonumber\\
\label{eq:fadia}
&&{}+\frac{2}{\ii a}L\left(v^>\right),
\end{eqnarray}
is the sum of $F^<_\zeta$ and of the $O(\ell^0)$ term of $F^>_\zeta$. This pure relaxation term differs from its steady-state limit (\ref{eq:selfss}) by an $O(1/t)$ correction, interpreted as a manifestation of the so-called ``afterglow" effect --- a distinctive feature of problems involving moving line sources: as time increases any point along the accelerated line receives spherical waves emitted in the past by increasingly distant points on the line, which leads to a tail in the local response.\cite{MORS53BART89} Expression (\ref{eq:fadia}) will be employed for analysis in the next Section.

What has just been said applies as well to $F_\alpha(t)$, the force with drag term included, provided one replaces everywhere $L(v)$ by $L_\alpha(v)$, and $p(v)$ by $p_\alpha(v)=\dd L_\alpha(v)/\dd v=p(v)+\ii w_0/\cS$. Only the $O(\ell^0)$ term in $F_\zeta^>$ is modified by this substitution: $m(v)$ is unchanged since the second derivative of $L_\alpha(v)$ does not depend on $\alpha$, and $F_\zeta^<$ is unchanged as well because terms linear in the velocity in $L$ cancel out in (\ref{eq:finfexpr}). Hence, the destruction of the ``old'' field is drag independent.

One could be tempted to use a low-order truncation of expansion (\ref{eq:fsupexp}) as a cheap alternative to (\ref{eq:sup}). However, there is a catch. The functions $L(v)$, $p(v)$ and $m(v)$ involve square roots, of the type $(1-v^2/\cS)^{1/2}$ for the screw dislocation, and $(1-v^2/\cS^2)^{1/2}$ and $(1-v^2/\cL^2)^{1/2}$ for the edge.\cite{PELL12} Thus, employing principal determinations, the functions $L(v)$, $p(v)$ and $m(v)$ of the complex variable $v$ have branch cuts on the (overlapping) semi-infinite intervals $|v|>\cS$ and $|v|>\cL$ of the real axis. In (\ref{eq:sup}), $\overline{v}$ is by definition of positive imaginary part, so that these branch cuts are never crossed: no discontinuity arises in the integrand, and (\ref{eq:sup}) causes no trouble. This not true any more with $v^>$ in (\ref{eq:vsup}) whose imaginary part $a/t-\dot{a}/2$ can be of any sign, implying the possibility of crossing branch cuts in expansion (\ref{eq:fsupexp}). Since physics requires continuity over time, we need to remove the branch-cut discontinuities of $L$, $p$ and $m$ by analytic continuation. This is achieved by gathering the Riemann sheets from the complete set of determinations of the square roots, which makes $L(v)$, $p(v)$ and $m(v)$ multivalued. Then, passing from one sheet to another in the course of evolution prevents a simple computation of $M^{(0)}(\dot{\zeta}^*,a/t)$, which turns out to be path-dependent in the complex plane. In our opinion, this complication disqualifies expansion (\ref{eq:fsupexp}) as an alternative to Eq.\ (\ref{eq:sup}).

\section{Numerical results}
\label{sec:numres}
\begin{figure}[!htbp]
\includegraphics[width=8.65cm]{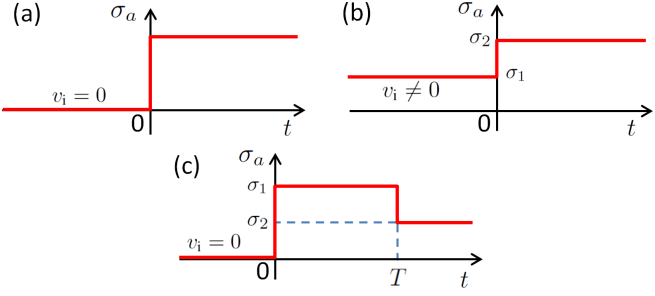}
\caption{\label{fig:fig2} (Color online) Loading types used. (a) single-step loading from rest; (b) single-step loading from steady-state with nonzero velocity; (c) double-step loading from rest.}
\end{figure}

This Section presents some numerical solutions of the EoM (\ref{eq:eomzetam12}) obtained with the algorithm described in Appendix \ref{sec:nummeth}. We apply it to an edge dislocation subjected to three different kinds of stress loadings (Fig.\ \ref{fig:fig2}).
In most of the calculations, a ratio $\cL/\cS=2$ is imposed between longitudinal and transverse wave speeds. This corresponds to a Poisson ratio $\nu=1/3$ typical of most metals, and to $\cR\simeq 0.93\,\cS$. For $\alpha=0$, the SS branch in Fig.\ \ref{fig:fig1} is degenerate at $\sigma_a=0$ with undetermined velocity.\cite{ROSA01} So, $\alpha\geq 10^{-4}$ is used hereafter. Our aim being to delineate some general features of the EoM, we allow for unrealistically high values of $\alpha$. The natural time unit is $\tau_0=d/\cS$, namely, the characteristic propagation time of a shear wave across the interplane distance.

\subsection{Single-step loading from rest, and dynamic subsonic-transonic transition}
\label{sec:sslsfr}
The first case of interest is that of \emph{single-step loading from rest}, where a stress $\sigma_a$ is instantaneously applied at $t=0$ and kept constant afterwards [Fig.\ \ref{fig:fig2}(a)].
\subsubsection{Overview}
\begin{figure}[!ht]
\includegraphics[width=7.2cm]{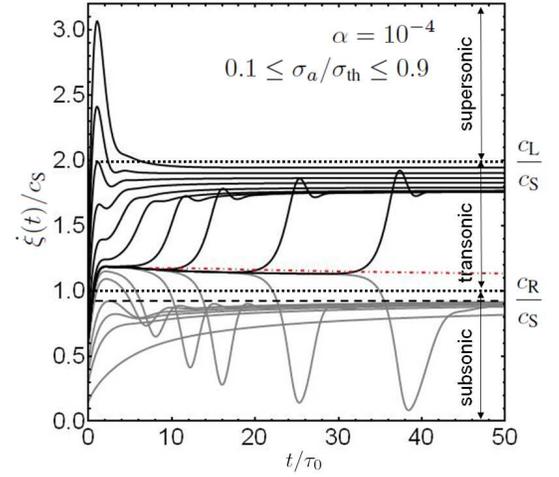}
\caption{\label{fig:fig3} (Color online) Single-step loading from rest. Velocity-time response $\dot{\xi}(t)$ of an edge dislocation subjected to increasing stress levels $\siga$ (from bottom to top). Dotted (resp., dashed) black horizontal lines: shear and longitudinal (resp., Rayleigh) wave speeds. Dot-dashed red line: approximation to the plateau state (see text).}
\end{figure}
\begin{figure}[!ht]
\includegraphics[width=7cm]{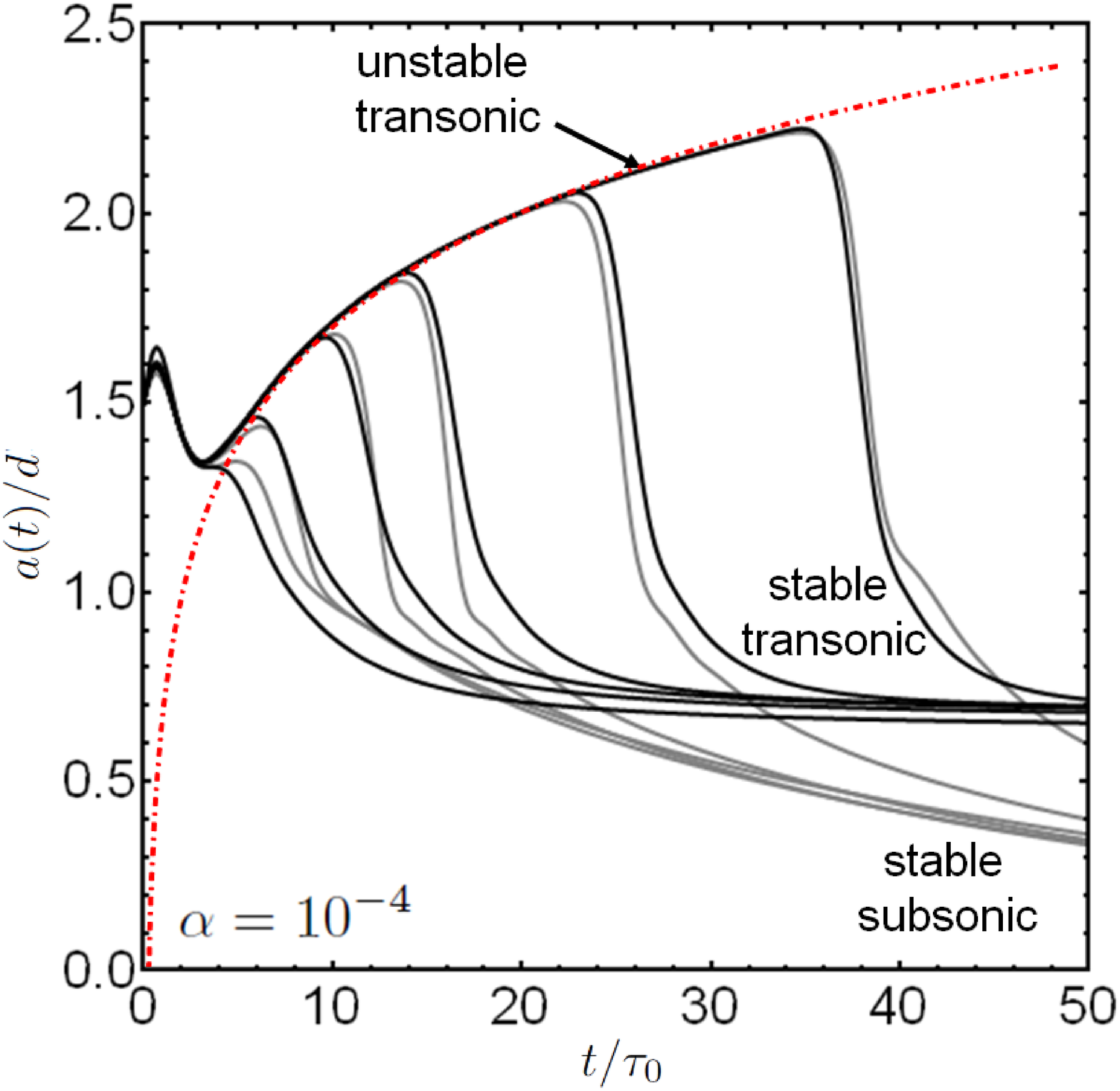}
\caption{\label{fig:fig4} (Color online) Core width-time response $a(t)$ of an edge dislocation subjected to instantaneously applied stresses $\siga$ close to $\sigma_c$. Dot-dashed: empirical function $y = 0.45 \ln(4.65 x)$, for comparison purposes.}
\end{figure}
\begin{figure}[!ht]
\includegraphics[width=6.5cm]{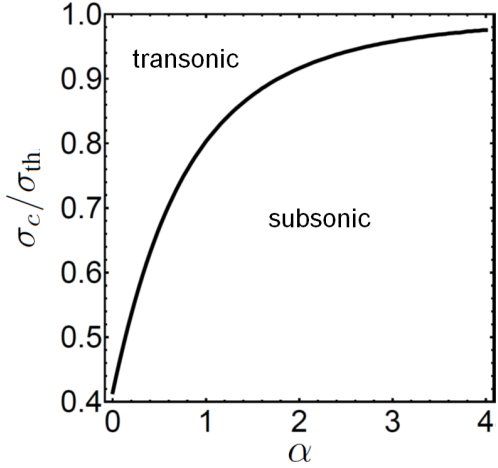}
\caption{\label{fig:fig5} Critical stress $\sigma_c$ vs.\ drag coefficient $\alpha$ for single-step loading from rest. The value at origin is $\sigma_c(\alpha=0)\simeq 0.415\,\sigth$.}
\end{figure}
\begin{figure*}[!ht]
\includegraphics[width=17.4cm]{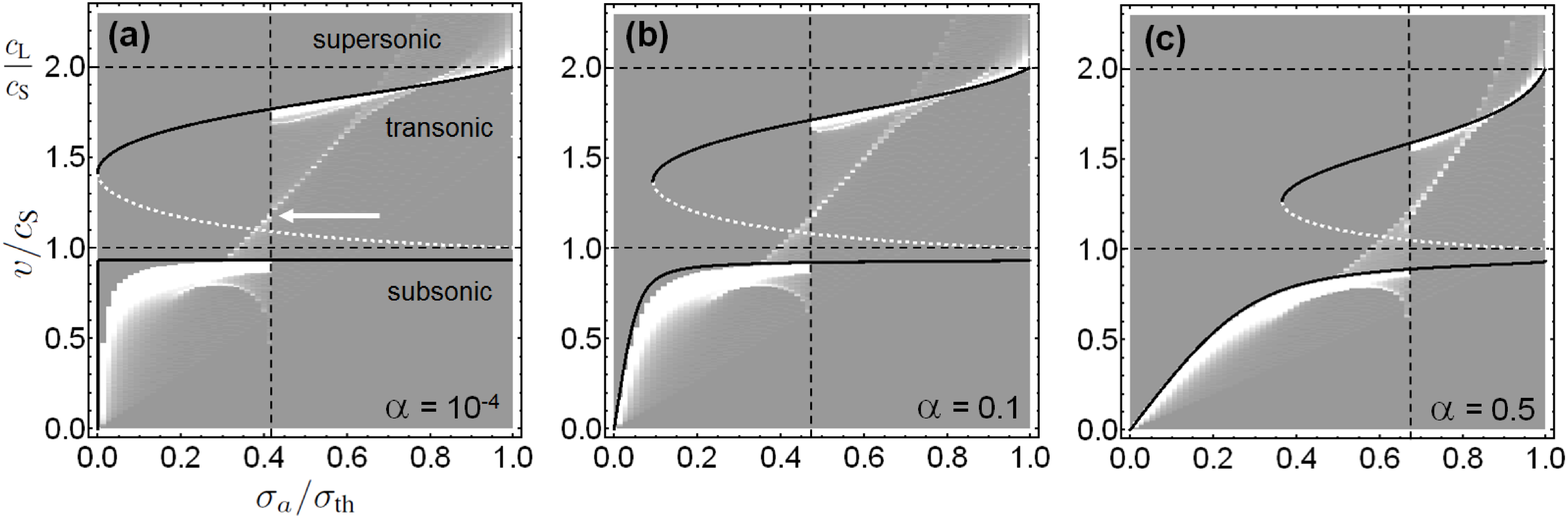}
\caption{\label{fig:fig6} Single-step loading from rest, for $t/\tau_0\leq 100$. From background grey (zero) to white (threshold), along the vertical axis: normalized distributions of non-uniform velocities $v=\dot{\xi}$ thresholded at $0.005$, vs.\ $\sigma_a$. Solid black (resp., dashed white): stable (resp., unstable) steady states of Fig.\ \ref{fig:fig1}. Vertical dashed line: critical stress $\sigma_c$. White arrow in (a): transient plateau state of Fig.\ \ref{fig:fig3}.}
\end{figure*}
Figure \ref{fig:fig3} represents velocity-time plots for $\alpha$ near zero, for increasing stresses in the range $0.1\leq\siga/\sigth\leq 0.9$ (from bottom to top), with emphasis around a special stress value to be analyzed shortly. After an initial velocity jump at $t=0^+$,\cite{ESHE53,PILL07} the dislocation accelerates smoothly over a time interval of order $\tau_0$. For low $\sigma_a$ motion is overdamped, whereas moderate stress results in damped core-induced velocity oscillations, akin to those observed in atomistic simulations by Olmsted \textit{et al.}\cite{OLMS05} In the initial acceleration stage, and for high applied stress, a velocity peak culminates in the transonic or even supersonic regimes, with no particular signal at wave-speed values. Past the initial stage subsequent evolution leads to either subsonic velocities bounded upwards by $\cR$, or transonic velocities, as given by the steady-state theory (Sec.\ref{sec:steady}). Both types of motions are separated by a \emph{dynamic critical stress} $\sigma_c(\alpha)$.

The closer $\sigma_a$ to $\sigma_c$, the longer the dislocation remains on an unstable transonic plateau of slowly decreasing velocity (Fig.\ \ref{fig:fig3}). On this plateau, the core width grows as $\ln t$ (Fig.\ \ref{fig:fig4}), before the dislocation either quickly leaps to a faster transonic state (black curves), or decays into the subsonic range (grey curves) depending on whether $\sigma_a\lessgtr\sigma_c$. In both cases, the core contracts rapidly during the transition. The explanation resides in that the expanding dislocation stores excess stacking-fault energy, to be released as it contracts, either in the form of a velocity boost, or in the form of elastic waves emitted as the shear Mach wavefront separates from the decelerating dislocation. In the latter case the dislocation may almost stop, in agreement with observations made on atomistic simulations,\cite{GUMB99,JING08} which has been interpreted as a consequence of the backwards momentum push of the detaching front.\cite{GUMB99}

The response at higher $\alpha$ values is similar, except that transients are further damped and that the asymptotic state is reached sooner (not shown). The threshold $\sigma_c(\alpha)$, represented in Fig.\ \ref{fig:fig5}, is determined by a dichotomy search from the asymptotic behavior. It is drag-dependent and defines a phase boundary in the $(\alpha,\sigma_a)$ plane.

The dynamics of the model is surveyed in Fig.\ \ref{fig:fig6} for three drag coefficients $\alpha=10^{-4}$ (a), $\alpha=0.1$ (b), and $\alpha=0.5$ (c), by comparing transient velocities to steady states. The figures display density maps of the velocities, obtained as follows. Calculations for 76 stress values $\sigma_a$ evenly spread in the interval $[0,\sigth]$, were run up to $t=100\,\tau_0$ to generate velocity curves such as in Fig.\ \ref{fig:fig3}. With time step $\delta t$ small enough to ensure good sampling, the obtained velocity sets $v_k=v(t=k\delta\tau)$ were distributed into 300 equispaced bins as a normalized frequency histogram.\cite{ROOS01} In the figures, one such distribution is plotted vertically in grey tones, for each $\sigma_a$. The whole array makes up the density map. For $\alpha$ small, because of slow relaxation, making meaningful comparisons with steady-state curves supposes long-time runs, which has the drawback of giving excessive weight to the vicinity of the asymptotic states. Therefore, to bring local extrema of the time-velocity curves into light, distributions are cut off at $0.005$, which is the value of the white regions of the maps, while the dominant grey tone represents value 0.

The spread of the white region in the subsonic range illustrates the slow character of relaxation towards asymptotic states. At low drag and low driving stress, the subsonic asymptotic state is far from being reached [Fig.\ \ref{fig:fig6}(a)], consistently with observations previously made on atomistic simulations.\cite{JING08} The lower envelope of the white regions bends downwards as the critical stress is approached, and represents the lowest velocities in Fig.\ \ref{fig:fig3}. The white line of transient states that ``connects" the subsonic and transonic stable branches represents the first local maximum of the velocity observed in Fig.\ \ref{fig:fig3}. Its presence indicates that $\siga$ is high enough to make core-induced effects noticeable.

\subsubsection{The transition as a delayed bifurcation}
\label{sec:delayedbif}
\begin{figure*}[!ht]
\includegraphics[width=12.6cm]{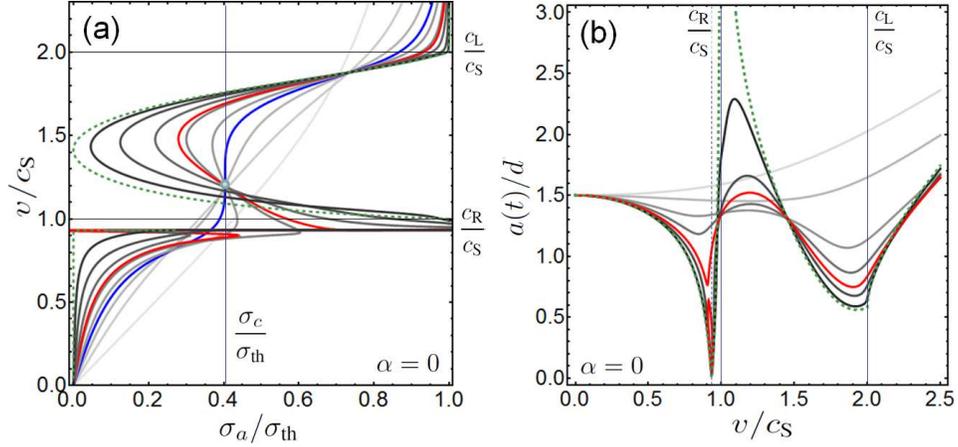}
\caption{\label{fig:fig7} (Color online) Simplified relaxation model, for $\cL=2\cS$ and $\alpha=10^{-4}$. (a) velocity/stress curves for times $t/\tau_0=0.5$, $1.5$, $2.5$, $4.3$, $6.25$, $10$, $20$, $60$ (lighter to darker grey), $t/\tau_0=3.4$  (blue) and $t/\tau_0=7$ (red). (b) Width $a(v,t)$ vs.\ velocity $v$ for $t/\tau_0=0.5$, $1$, $2.5$, $4$, $10$, $40$  (lighter to darker grey) and $t/\tau_0=6.25$  (red).  In (a) and (b), the dashed curve (green) is the steady-state solution.}
\end{figure*}
To study the nature of the transition, let us briefly leave the full EoM. Here, we simply enrich the steady-state equation by using, in a slightly modified form, the `adiabatic' contribution to the self-force isolated in Sec.\ \ref{sec:local}. Thus, we substitute to Eq.\ (\ref{eq:steadyf}) the equation
\begin{subequations}
\label{eq:relaxationmodel}
\begin{equation}
\label{eq:relaxadia}
\widetilde{F}^{\text{adia}}_\alpha(v^>,v^<)=-\ii\,b\,\sigma_{\rm th}g\left(\sigma_a/\sigma_{\rm th}\right),
\end{equation}
where
\begin{eqnarray}
\widetilde{F}^{\text{adia}}_\alpha&=&\frac{2}{t}\left[
p\left(v^<\right)-p_\alpha\left(v^>\right)-\frac{L(v^<)-L(\vi+\ii 0^+)}{v^<-\vi}\right]\nonumber\\
&&{}+\frac{2}{\ii a}L_\alpha\left(v^>\right),\\
v^<&=&v+\ii(a+\ai)/(2t),\\
v^>&=&v+\ii a/t.
\end{eqnarray}
\end{subequations}
The above force $\widetilde{F}^{\text{adia}}_\alpha$ was obtained from empirically modifying $F^{\text{adia}}_\zeta$ in (\ref{eq:fadia}) in several ways, because $F^{\text{adia}}_\zeta$ is not much helpful as it stands. First, drag was added to $F^>$ according to the remark closing Sec.\ \ref{sec:local}; next, expression (\ref{eq:vinf}) of $v^<$ was simplified by means of the approximation $\xi(t)/t\simeq v$, which neglects accelerations; finally, expression (\ref{eq:vsup}) of $v^>$ was simplified by taking $\dot{\zeta}^*(t)\simeq v$, which neglects $\dot{a}$. Equations (\ref{eq:relaxationmodel}) are solved for $a(t)$ and $v(t)$ as functions of $t$ and $\sigma_a$.

Results are shown in Figs.\ \ref{fig:fig7}. By construction, the steady-state velocity branches are retrieved for $t=+\infty$, and Fig.\ \ref{fig:fig7}(a) depicts how they form. Keeping $\siga$ fixed, the solution $v(\sigma_a,t)$ of Eqs.\ (\ref{eq:relaxationmodel}) is unique at small times. However, branch separation takes place near $t=3.4\tau_0$. Afterwards, the velocity either turns subsonic or transonic depending on $\sigma_a\lessgtr 0.41\sigth$. So, this approximation captures the correct stress threshold. The ``turning point'' marked out by a dot in Fig.\ \ref{fig:fig7}(a) slowly moves down towards the unstable steady-state branch (dashed). Its evolution with $t$ is represented as the dot-dashed line in Fig.\ \ref{fig:fig3}, showing that it lies at the origin of the unstable plateau. Parenthetically, we must indicate that ---as useful as it is--- approximation (\ref{eq:relaxationmodel}) has pathologies. For instance, the solutions for $a(t)$ in  Fig.\ \ref{fig:fig7}(b) have several branches near $v\simeq\cR$ at small times, that coalesce just after $t\simeq 6.25\tau_0$.

The S-shape of the curves in Figs.\ \ref{fig:fig7}(a) is typical of an imperfect Hopf bifurcation\cite{STRO00} in which both the time and $\siga$ act as control parameters. We note  [Fig.\ \ref{fig:fig7}(b)] that near to bifurcation time $t=3.4\,\tau_0$, at the bifurcation velocity $v/\cS\simeq 1.2$, $\dot{a}$ vanishes, consistently with our approximations. However, Eqs.\ (\ref{eq:relaxationmodel}) cannot reproduce the tunable delay observed in Fig.\ \ref{fig:fig3}. Therefore, this delay must be caused by the inertial terms that we neglected. The following picture emerges: so to speak, the transition involves a slow, relaxation-like, process that shapes the stage on which a faster but inertia-controlled evolution takes place. Such \emph{delayed bifurcations},\cite{MARE96} rather common in physical and biological sciences, are the subject of intense research\cite{ZHEN12} subtended by a sophisticated mathematical theory.\cite{NEIS87} Hereafter, we limit ourselves to a few basic observations, leaving formal characterizations to the future.
\subsubsection{Delay to bifurcation and Lyapunov exponent}
\begin{figure*}[!ht]
\includegraphics[width=12.6cm]{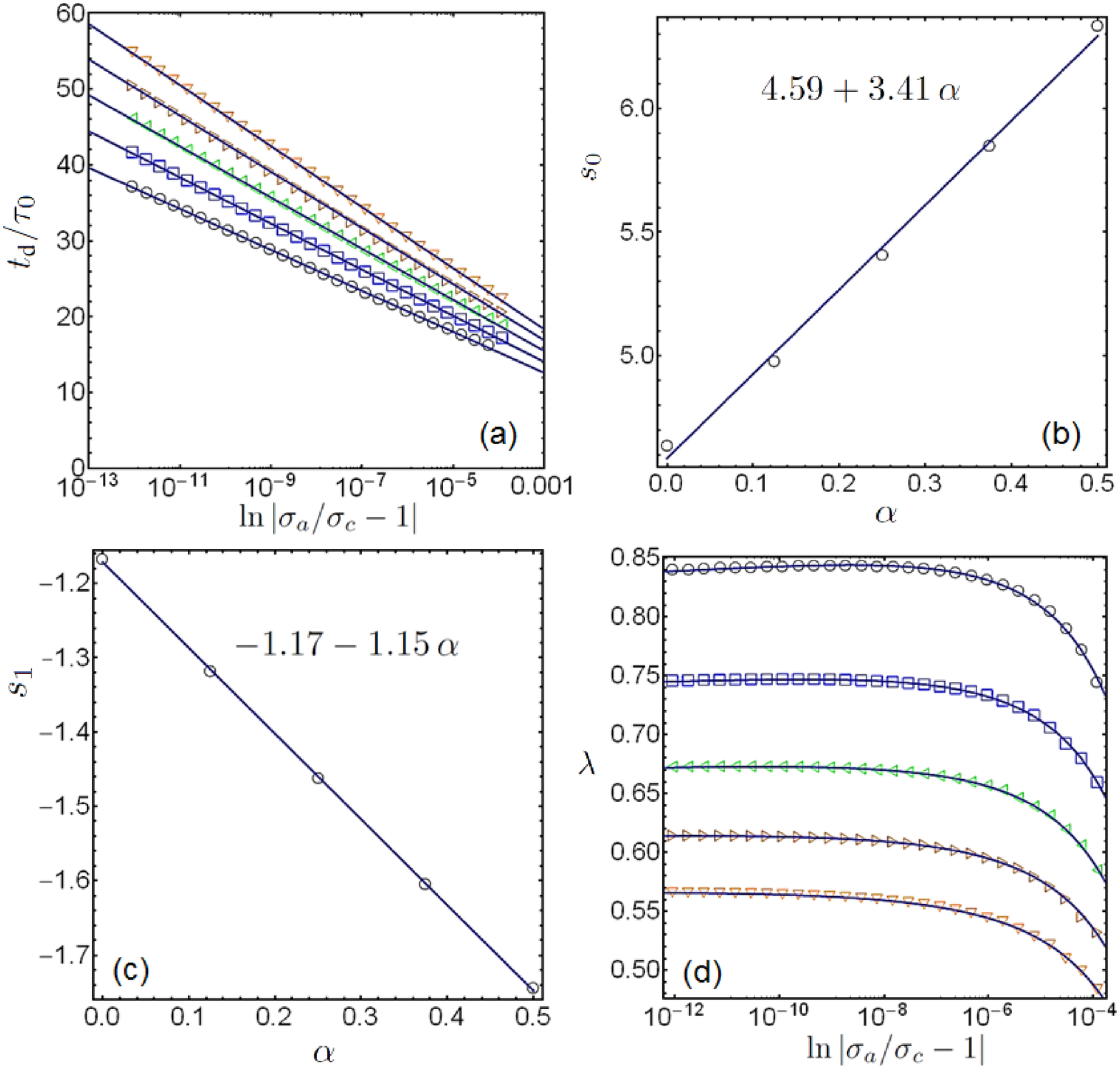}
\caption{\label{fig:fig8} (Color online) Analysis of the delay to bifurcation (``plateau length"). Solid lines represent fits. a) Delay $t_{\text{d}}(\sigma_a,\alpha)$ vs.\ applied stress $\sigma_a$ for drag coefficients $\alpha=0$, $0.125$, $0.25$, $0.375$ and $0.5$ (from bottom to top) and fits by Eq.\ (\ref{eq:tdfit}). b) Intercept $s_0(\alpha)$ and c) slope $s_1(\alpha)$ of the fits in (a). d) Lyapunov exponent $\lambda(\sigma_a,\alpha)$ vs.\ applied stress for the values of $\alpha$ in (a), and fits by Eq.\ (\ref{eq:lambdafit}).}
\end{figure*}

Returning to the full EoM, the delay time to bifurcation, denoted by $t_{\text{d}}(\sigma_a,\alpha)$, is extracted as follows. We first estimate the threshold $\sigma_c(\alpha)$ with 15-digits accuracy for several values of $\alpha$. Setting $\sigma_a$ equal to this estimate yields a reference velocity curve $v_{\text{ref}}(t;\sigma_c,\alpha)$ with a plateau longer than $40\,\tau_0$, sufficient for our purpose. Next, the difference $\Delta v(t;\sigma_a,\alpha)=v(t;\sigma_a,\alpha)-v_{\text{ref.}}(t;\sigma_c,\alpha)$ is computed for applied stresses $\sigma_a=\sigma_c(1\pm 2^{-i})$ close enough to $\sigma_c$ (with $i=13,\ldots,40$), which produces plots of $\Delta v$ with longer and longer delays before ``lift-off". To isolate the latter, assumed to be of exponential form (see Fig.\ \ref{fig:fig3}), time evolution is stopped at the first inflexion of $\Delta v(t)$. The resulting data sets are fitted to the form
\begin{equation}
\label{eq:expfit}
\Delta v(t;\sigma_a,\alpha)\simeq \cS\, \exp\left\{\lambda(\sigma_a,\alpha)\left[t-t_{\text{d}}(\sigma_a,\alpha)\right]/\tau_0\right\},
\end{equation}
where the Lyapunov exponent, $\lambda$, and $t_d$ are fitting parameters. The resulting fits to (\ref{eq:expfit}) are indistinguishable to the eye from the data. We find [Fig.\ \ref{fig:fig8}(a)] that $t_d$ has \emph{logarithmic dependence in the distance to the critical stress}, namely,
\begin{equation}
\label{eq:tdfit}
t_{\text{d}}(\sigma_a,\alpha)\simeq s_0(\alpha)+s_1(\alpha)\ln\varepsilon(\sigma_a,\alpha),
\end{equation}
where $\varepsilon(\sigma_a,\alpha)=|\sigma_a/\sigma_c(\alpha)-1|$. Both coefficients $s_0(\alpha)$  [Fig.\ \ref{fig:fig8}(b)] and  $s_1(\alpha)$  [Fig.\ \ref{fig:fig8}(c)] depends quasi-linearly on $\alpha$ in the considered range. Moreover, $\lambda(\sigma_a,\alpha)$ tends to a constant as $\sigma_a$ approaches $\sigma_c(\alpha)$, and decreases away from it with logarithmic corrections well-represented by the three-parameter expression
\begin{equation}
\label{eq:lambdafit}
\lambda(\sigma_a,\alpha)\simeq \sum_{k=0}^2 A_k(\alpha)[\ln\varepsilon(\sigma_a,\alpha)]^{-k},
\end{equation}
where $A_0=\lambda(\sigma_c,\alpha)>0$ is of order one, and $A_1$ and $A_2$ are negative of order $1$ and $10$, respectively [Fig.\ \ref{fig:fig8}(d)]. All three parameters increase mildly in a non-linear way with $\alpha$. Given the arbitrariness of (\ref{eq:lambdafit}), this point is not elaborated further. The dislocation is expelled from the plateau with greater strength near the critical line, where exponent $\lambda$ is the largest (see Fig.\ \ref{fig:fig3}). The results of Fig.\ \ref{fig:fig8}, obtained for $\sigma_a$ very close to $\sigma_c$, do not depend on which side of $\sigma_c$ the applied stress is. The values indicated, computed with algorithmic time step $\delta t=\tau_0/20$ (see Appendix B), slightly vary with $\delta t$ while the fitting forms remain valid.

\subsection{Single-step loading from nonzero velocity}
\label{sec:sslnzv}
\begin{figure}[!htbp]
\includegraphics[width=8.65cm]{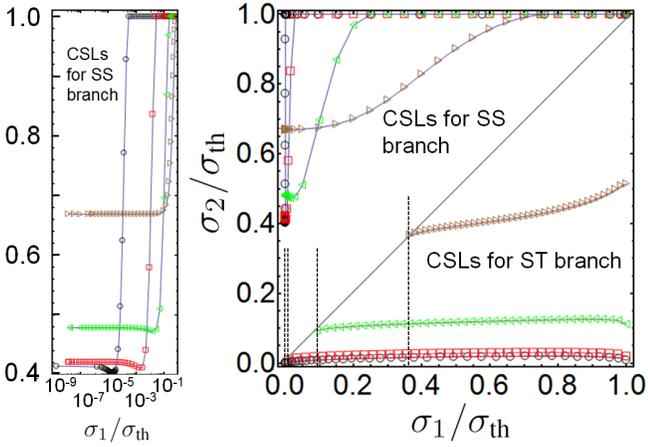}
\caption{\label{fig:fig9} (Color online) Single-step loading from steady state with nonzero velocity. For various values of $\alpha$ are represented drag-dependent critical stress lines (CSL) in the plane $(\sigma_1,\sigma_2)$. Pairs $(\sigma_1,\sigma_2)$ lying \emph{above}  (resp.\ {\emph below}) one particular CSL lead as $t\to\infty$ to an asymptotic velocity state of the ST (resp.\ SS) type. Both cases where the initial velocity $\vi(\sigma_1)$ lies either on the SS branch or on the ST branch (see Fig.\ \ref{fig:fig1}) are examined.  Left: small-$\sigma_1$ blow-up of the main plot. CSLs are sampled as circles (black, $\alpha=10^{-4}$), squares (red, $\alpha=10^{-2}$), left triangles (green, $\alpha=0.1$), or right triangles (brown, $\alpha=0.5$). For CSLs relative to the ST branch of initial velocities, only pairs $(\sigma_1,\sigma_2)$ where $\sigma_1$ lies at the right of the corresponding vertical dashed line (that must be extended up to $\sigma_2=1$) are relevant.}
\end{figure}
We next examine the effect of single-step loading from $\sigma_a=\sigma_1$ to $\sigma_2$, the dislocation being now at negative times in an initial state of \emph{finite} steady velocity $\vi=v(\sigma_1) > 0$ [Fig.\ \ref{fig:fig2}(b)]. Two possibilities arise, since this initial state can belong either to the SS branch or---for $\sigma_a$ large enough if $\alpha>0$---to the ST branch of Fig.\ \ref{fig:fig1}.

Figure \ref{fig:fig9} represents for some values of $\alpha$ the critical stress line (CSL) that separates, in the plane $(\sigma_1,\sigma_2)$, loading conditions leading to subsonic asymptotic states from those leading to transonic ones. The dislocation has constant velocity on the diagonal where $\sigma_2=\sigma_1$. Achieving an upwards (resp.\ downwards) velocity shift requires increasing (resp.\ decreasing) the applied stress. Hence, subsonic-to-transonic transitions can only occur within a subregion of the domain $\sigma_2>\sigma_1$, whereas transonic-to-subsonic transitions can only occur within the complementary domain $\sigma_2<\sigma_1$. Within either of these domains, a point $(\sigma_1,\sigma_2)$ lying below (resp., above) any given CSL leads to a subsonic (resp., transonic) steady state.

The figure shows that initial steady states of nonzero velocity have huge inertia, which we attribute to their ``field-dragging" character, and that dissipation proves essential in helping transitions to take place. Indeed, consider the domain $\sigma_2>\sigma_1$. When $\alpha\to 0$, we see that subsonic-to-transonic transitions are impossible for finite initial velocity $\vi>0$ (i.e.\ finite applied stress $\sigma_1>0$). However, they become allowed when $\alpha\not=0$, in the low-stress interval proportional to $\alpha$ where (roughly) the linear drag-controlled mobility law $v\simeq \widetilde{B}\sigma_a$ predominates. On the other hand, transonic-to-subsonic decays in the domain $\sigma_2<\sigma_1$ are possible \emph{whatever $\alpha$}, even though for $\alpha\to 0$ the stress $\sigma_2$ must be lowered below $0.025\sigth$ for them to occur. In this domain as well, the figure shows that a finite $\alpha$ eases transitions by shifting upwards the CSL. Thus, transonic-to-subsonic transitions when $\alpha\to 0$ should be attributed to the $\alpha$-independent Mach-cone dissipation in the initial transonic state.

\subsection{Double-step loading from rest}
\label{sec:dssr}
\begin{figure}[!htbp]
\includegraphics[width=6.5cm]{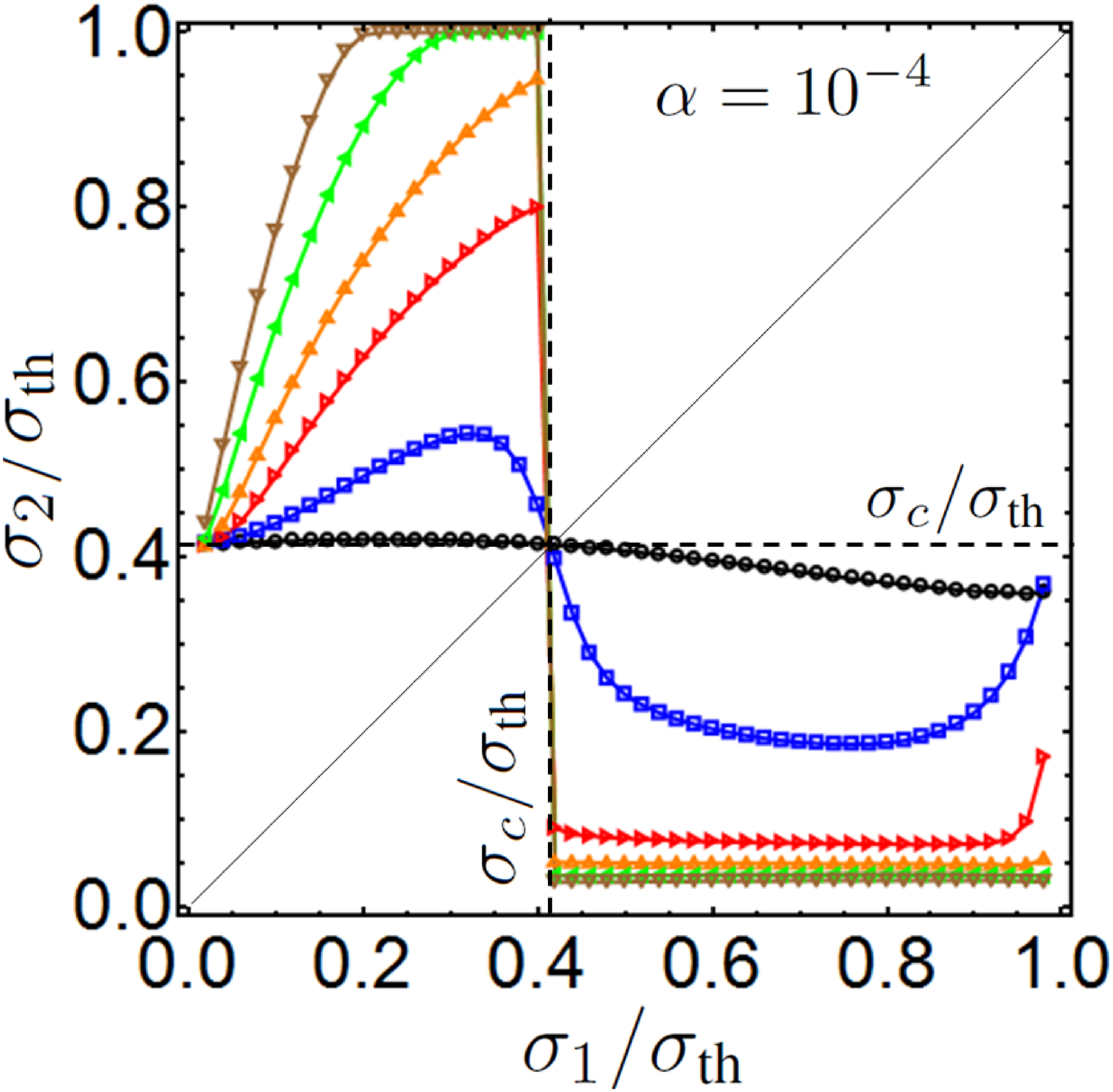}
\caption{\label{fig:fig10} (Color online) Critical stress lines (CSLs) for double-step loading from rest. Loading $\sigma_1$ (respectively, $\sigma_2$) is applied at $t=0$ (respectively, $t=T>0$). The dislocation is asymptotically subsonic below any CSL, and asymptotically transonic above it. Circles (black), CSL for $T=\tau_0$; squares (blue), $T=5\,\tau_0$; right triangles (red), $T=15\,\tau_0$; up triangles (orange), $T=30\,\tau_0$; left triangles (green), $T=60\,\tau_0$; down triangles (brown), $T=100\,\tau_0$.}
\end{figure}
\begin{figure*}[!ht]
\includegraphics[width=17.4cm]{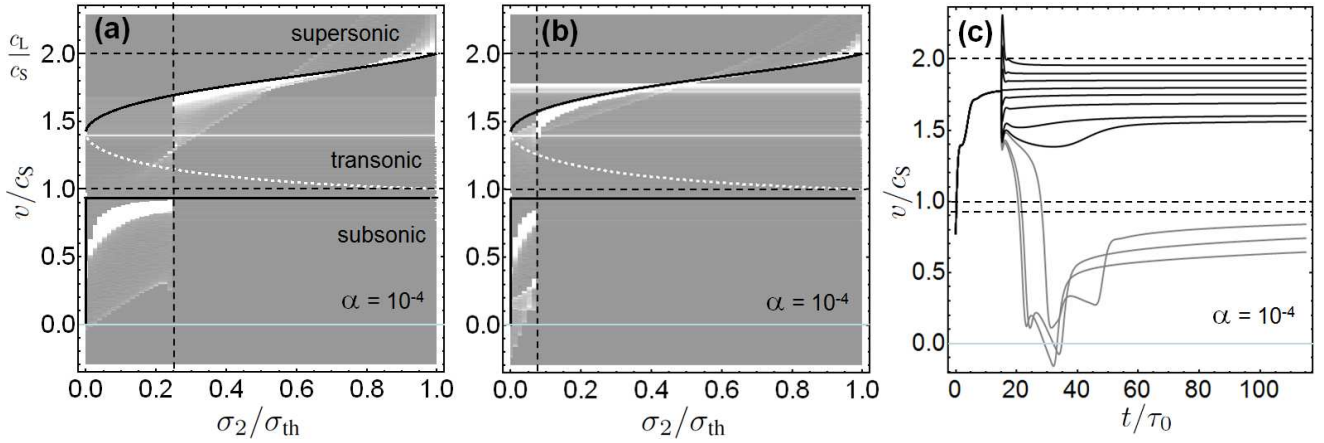}
\caption{\label{fig:fig11} (a) and (b) Velocity distributions for double-step loading from rest, with $\sigma_1/\sigth=0.5$ in the first stage. Second stage occurs at $T=5\tau_0$ in (a), and $T=15\tau_0$ in (b). The time window is $t\leq 100\tau_0+T$. In (c), velocity vs.\ time plots corresponding to case (b), for the following selected values of $\sigma_2/\sigth$ (from bottom to top): $0.013$, $0.026$, $0.067$ (grey), $0.080$, $0.120$, $0.253$, $0.387$, $0.507$, $0.653$, $0.787$, and $0.920$ (black).}
\end{figure*}
We finally consider double-step loading from rest, which has been employed in atomistic simulations\cite{GUMB99} to ``lock" the dislocation onto small-stress transonic states that were unavailable with single-step loading. Our EoM reproduces this effect, for which no explanation has been given so far. The dislocation being initially \emph{at rest} under zero stress, stress $\sigma_1$ is applied from $t=0$ to time $t=T>0$, and stress $\sigma_2$ is applied next at $t=T$ and kept constant thereafter [Fig.\ \ref{fig:fig1}(c)]. The resulting CSLs are represented in Fig.\ \ref{fig:fig10} for $\alpha=10^{-4}$ and increasing values of $T$.

Stress $\sigma_1$ is irrelevant in the limit $T\to 0$ where the CSL becomes the horizontal line $\sigma_2=\sigma_c(\alpha)$---the single-step-loading critical stress discussed in Sec.\ \ref{sec:sslsfr}. In the opposite limit $T\to\infty$ the dislocation has enough time to relax towards the asymptotic state determined by $\sigma_1$, prior to being subjected to $\sigma_2$, so that the situation approaches that of Fig.\ \ref{fig:fig1}(c) examined in Sec.\ \ref{sec:sslnzv}. However, the branch-selection process described in Sec.\ \ref{sec:sslsfr} now takes place after the first acceleration step. Consequently, the trend observed in Fig.\ \ref{fig:fig10} is that the $T=\infty$ limit of the CSLs stands as a discontinuous combination of the CSLs of Fig.\ \ref{fig:fig9}, the chosen one depending on whether $\sigma_1\lessgtr \sigma_c(\alpha)$. Moreover, the slow saturation with $T$ of the CSLs in the low-$\sigma_1$ region of Fig.\ \ref{fig:fig10} makes clear that the CSLs of Fig.\ \ref{fig:fig9} are only ideal ones, never observed at finite times. The reason is that in an infinite medium the build-up of a steady elastic field configuration takes infinite time.

Figure \ref{fig:fig10} also explains how to reach steady transonic states at small applied stress: from rest, apply a stress $\sigma_1>\sigma_c(\alpha)$ to drive the dislocation into the transonic regime, wait for some time $T$, and then decrease the stress level to a value $\sigma_2$ slightly above the CSL at this $T$. This recipe is illustrated by Fig.\ \ref{fig:fig11} where $\sigma_1=0.5>\sigma_c(\alpha=10^{-4})\simeq 0.41$. The larger $T$, the smaller $\sigma_2$ can be, and the closer the asymptotic state to the left tip of the ST branch. Consistently with Fig.\ \ref{fig:fig10}, the critical stress for $\sigma_2$ is near to $0.25\sigth$ in (a) and to $0.075\sigth$ in (b).

For $\alpha\simeq 0$ this point is the radiation-free transonic state discovered by Eshelby, where in principle the dislocation can move at zero applied stress without dissipation. Figures \ref{fig:fig9} and \ref{fig:fig10} indicate that it cannot be reached dynamically, as for $\sigma_1>\sigma_c(\alpha)$, any stress $\sigma_2$ below the ``black/circle" CSL in Fig.\ \ref{fig:fig9} inevitably leads to decay into the subsonic range. According to Sec.\ \ref{sec:sslnzv}, this is caused by Mach-cone dissipation. We infer that dissipation destabilizes the dislocation before it has any chance to settle in the dissipation-free state. The drag $\alpha$ being finite is another reason why in practice such a dissipation-free state is unreachable.

Finally, Fig.\ \ref{fig:fig11}(c) presents velocity-time curves for some of the stresses $\sigma_2$ used in Fig.\ \ref{fig:fig11}(b), at same $\sigma_1$. For a huge stress drop, the dislocation can temporarily recoil under the backwards push of the detaching Mach cone. Subsequent velocity bounces, attributable to core-width dynamics, are observed before the dislocation resumes a regular forward motion.

\subsection{Comparisons with atomistic data}
\begin{figure}[!ht]
\includegraphics[width=7.20cm]{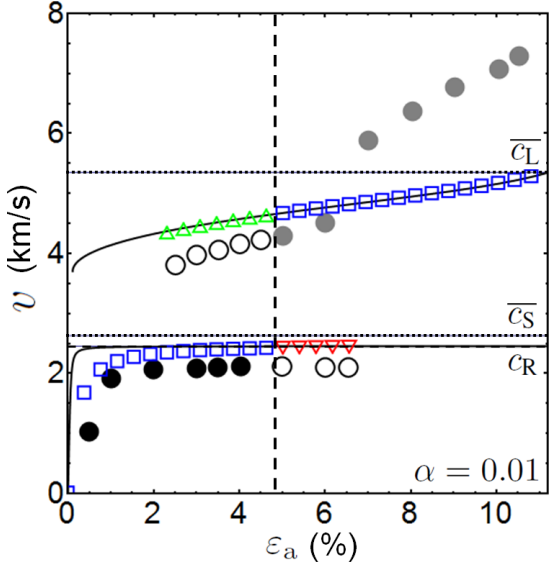}
\caption{\label{fig:fig12} (Color online) Comparisons with atomistic simulation data on tungsten by Jin, Gumbsch and Gao (Fig.\ 2 of Ref.\ \onlinecite{JING08}). Full and open disk symbols represent dislocation velocities in the twinning direction of motion, achieved in Ref.\ \onlinecite{JING08} with single-step loading (black and grey disks) and smoothed double-step loading (open circles). Other symbol shapes represent results at $t=10$ ps from the present model under single-step (square symbols) and double-step (up and down triangles) loading conditions.}
\end{figure}
We compare in Fig.\ \ref{fig:fig12} the model with recent atomistic simulation data by Jin, Gumbsch and Gao,\cite{JING08} obtained in tungsten with a Finnis-Sinclair potential.\cite{FINN84} Because tungsten is a quasi-isotropic bcc metal, those simulations constitute the most appropriate benchmark for the model. To cope with the nonlinear elastic effects at high applied shear distortion $\varepsilon_{\rm a}$ pointed out in Ref.\ \onlinecite{JING08}, which result in strain-dependent wave speeds, we use average speeds defined as $\overline{c}_{{\rm S},{\rm L}}\equiv \varepsilon_{\rm th}^{-1}\int_0^{\varepsilon_{\rm th}}\dd\varepsilon\,c_{{\rm S},{\rm L}}(\varepsilon)$ where the integral of the wave speed curves $c_{{\rm S},{\rm L}}(\varepsilon)$ of Ref.\ \onlinecite{JING08} is carried out up to the critical distortion $\varepsilon_{\rm th}=0.115$ above which the simulated crystal is unable to sustain a rigid shear.\cite{JING08} We identify the corresponding critical shear stress $\sigth=14.5$ GPa\cite{JING08} with the theoretical shear stress of the Peierls model. We obtain $\overline{c}_{\rm S}=2629$ m/s and $\overline{c}_{\rm L}=5350$ m/s. Using the density $\rho=19.257$ g/cm${}^3$, we next define over the considered strain range a consistent effective shear modulus $\overline{\mu}\equiv\rho\,\overline{c}_{\rm S}^2=134.61$ GPa, an effective interplane distance $\overline{d}\equiv\overline{\mu}b/(2\pi\sigth)=3.89$ \AA, and an effective characteristic time $\tau_0\equiv \overline{d}/\overline{c}_{\rm S}\simeq 0.148$ ps. We note that $\overline{d}\simeq 3.01 d$ is somewhat larger than the crystallographic interplane distance $d=a_0/\sqrt{6}$, where $a_0=3.165$ {\AA} is the lattice parameter, relevant to the $[112]$ glide plane in the atomistic simulation.\cite{JING08} This discrepancy is due to the oversimplified cohesive-zone approximation in the Peierls model.

In bcc metals motion is asymmetric with respect to the twinning and anti-twinning directions.\cite{JING08} Since the model cannot account for such an asymmetry, we limit our comparisons to motion in the anti-twinning direction. We display velocities values at 10 ps, the typical duration of the simulations.\cite{JING08} In Fig.\ \ref{fig:fig12}, black and grey disks are atomistic simulation data for steady-state velocities obtained with single-step loading. Remarkably, taking $\alpha=0.01$---a value consistent with typical values for metals (see Sec.\ \ref{sec:steady})---makes the critical distortion for the subsonic/transonic transition in single-step loading (square symbols) coincide with that of atomistic data, as shown in Fig.\ \ref{fig:fig12}. Moreover, atomistic data obtained with a smoothed kind of double-step loading\cite{JING08} (open symbols) are reasonably reproduced by employing abrupt double-step loading with either $\sigma_1=0.5\,\sigth$ for $0\leq t\leq 6.3_,\tau_0\simeq 0.93$ ps (up triangles), or $\sigma_1=0.3\,\sigth$ for $0\leq t\leq 7\,\tau_0\simeq 1.04$ ps (down triangles), and $\sigma_2=\overline{\mu}\varepsilon_a$ during the rest of the calculation.

\section{Concluding discussion}
\label{sec:concl}
Thus, the proposed EoM allows one to pinpoint for the first time the existence of a well-defined drag-dependent threshold stress in the subsonic/transonic transition of edge dislocations, which we identified as a delayed bifurcation. The present work demonstrates the need to take into account core-width variations in dislocation motion, which prove crucial to the transition examined. Confirmation is provided by a reinterpretation of simulation data that departs from a previous attempt.\cite{ROSA01} In particular, we can now closely approach these data by using a realistic value of the phenomenological phonon drag coefficient, namely, $\alpha\sim 10^{-2}$. Moreover, critical stress lines such as obtained in Sec.\ \ref{sec:dssr}, easy to compute with the model, might be used as guidelines for future atomistic simulations. For instance, it would be interesting to reproduce the recoil of the dislocation computed in Fig.\ \ref{fig:fig11}(c). By contrast, long unstable plateau states of duration $\propto\ln|\siga-\sigma_{\rm c}|$ should not be observable, for their obtention requires a very accurate determination of the critical threshold, which thermal fluctuations will most probably forbid. Whereas the present theory is essentially valid for an infinite medium, it is relevant to finite systems as well inasmuch as wave reflection on boundaries can be ignored (time of flight small enough, or quasi-perfect absorbing boundary conditions).

Figure \ref{fig:fig12} shows that discrepancies with simulations remain. First, the subsonic and transonic branches of the model lie above simulation data. While a better treatment of nonlinear elasticity could somewhat reduce the mismatch, the differences involved more probably indicate that we underestimate radiative drag. Additional possibly relevant sources of radiative drag could be investigated by slight modifications of the present framework, such as the periodic oscillations of the dislocation on the Peierls potential,\cite{ALSH71,KOIZ01} and the extension of the core normal to the glide plane, which cannot be excluded.\cite{GUMB99} Another limitation of the present work is that the parametrization employed imposes a symmetric core shape. One could instead expect the core to become asymmetric in transient regimes, due to motion-induced forwards-backwards symmetry breaking. Also, alleviating the absence of \emph{steady} supersonic steady states at stresses lower than $\sigth$ in the model would probably require enriching the model with lattice dispersion effects.\cite{ESHE56,ROSA01,MARI06} This is challenging bacause the collective-variable approach relies on the existence of explicit and reasonably easy-to-handle steady-state solutions. However, it should be noted that, due to the instability of straight dislocation lines at high velocities,\cite{MARI04} endowing the model with steady supersonic states might not be much relevant to `real' (i.e.\ non-rectilinear) dislocations. Besides, whereas transient supersonic states have been observed in a two-dimensional plasma crystal, recent measurements\cite{NOSE07} did not provide conclusive evidence for steady supersonic states in such systems.

The present CV formalism could easily lend itself to the use of a mul\-ti-\-dis\-lo\-ca\-tion ansatz, or to more elaborate parametrizations, by means of which other types of solutions could be explored. This includes twinning dislocations, or even plane sets of dislocations with Burgers charge $\pm b$ in dynamic interaction, which would constitute one further step towards fully dynamic two-dimensional DD simulations. Another immediate perspective consists in extending the present EoM to anisotropic elasticity,\cite{HIRT82} to exploit the wealth of available simulation data on anisotropic materials.\cite{OLMS05,MARI06,TSUZ08,GILB11}.

\acknowledgments
The author thanks Y.\ Ben-Zion and A.\ Le Pichon for useful remarks concerning the ``local" loss term in crack theory, and bound\-ary-int\-eg\-ral-equ\-a\-tion methods, respectively, and E.\ Bitzek, C. Denoual and D.\ Mordehai for stimulating discussions. Thanks are also due to anonymous reviewers for useful suggestions and remarks.
\appendix
\section{Explicit constrained equations}
\label{sec:evoleq}
For completeness, we derive here the constrained evolution equations for $\Delta\eta(x,t)$, $a(t)$ and $\xi(t)$ in full form, according to the method outlined in Sec.\ \ref{sec:collvars}. Consider first Eq.\ (\ref{eq:equsmotde}). Introducing $C=(1+\alpha)\mu/(2\cS)$, and omitting $x$ and $t$, we write $\mathcal{F}$ as ($[\cdot]$ indicates a functional dependency)
\begin{equation}
\mathcal{F}[\eta]=\sigma_\eta[\eta]-C\partial_t\widetilde{\eta}+\sigma_a-f'(\eta),
\end{equation}
where $\sigma_\eta[\eta]$ stands in this section for expression (\ref{eq:pnstress}) amputated from the ``local" loss term proportional to $\partial_t\widetilde{\eta}$. Taking $\eta=\eta_0+\Delta\eta$, so that $\widetilde{\eta}=\eta_0-\eta^{\rm e}+\Delta\eta$, one has
\begin{equation}
\partial_t\widetilde{\eta}=\partial_t\widetilde{\eta}_0+\partial_t\Delta\eta,
\end{equation}
where $\partial_t\widetilde{\eta}_0=-\rho_1(\dot{a}/a)-\rho_0 \dot{\xi}$. Moreover, from Eq.\ (\ref{eq:dconstr}) follows that
\begin{subequations}
\begin{eqnarray}
\frac{\dot{a}}{a}&=&\frac{1}{a}
\frac{
\braket{\partial_\xi\rho_0}{\Delta\eta}
\braket{\rho_1}{\partial_t\Delta\eta}-
\braket{\partial_\xi\rho_1}{\Delta\eta}
\braket{\rho_0}{\partial_t\Delta\eta}
}
{
\braket{\partial_a\rho_0}{\Delta\eta}
\braket{\partial_\xi\rho_1}{\Delta\eta}-
\braket{\partial_a\rho_1}{\Delta\eta}
\braket{\partial_\xi\rho_0}{\Delta\eta}
}\nonumber\\
&\equiv& A[\Delta\eta],\\
\dot{\xi}&=&
\frac{
\braket{\partial_a\rho_1}{\Delta\eta}
\braket{\rho_0}{\partial_t\Delta\eta}-
\braket{\partial_a\rho_0}{\Delta\eta}
\braket{\rho_1}{\partial_t\Delta\eta}
}
{
\braket{\partial_a\rho_0}{\Delta\eta}
\braket{\partial_\xi\rho_1}{\Delta\eta}-
\braket{\partial_a\rho_1}{\Delta\eta}
\braket{\partial_\xi\rho_0}{\Delta\eta}
}\nonumber\\
&\equiv& X[\Delta\eta],
\end{eqnarray}
\end{subequations}
which introduces the functionals $A$ and $X$ of $\Delta\eta$. They are of zero degree of homogeneity in this quantity, and also depend of $a$ and $\xi$ (dependence omitted). Thus the equation $\mathcal{F}[\eta_0+\Delta\eta]=0$ becomes the following equation for $\Delta\eta$:
\begin{eqnarray}
&&\sigma_\eta[\Delta\eta]-C\left\{\partial_t\Delta\eta-\rho_1 A[\Delta\eta]-\rho_0 X[\Delta\eta]\right\}\nonumber\\
&&\qquad{}+\sigma_a-f'(\eta_0+\Delta\eta)+\sigma_\eta[\eta_0]=0,
\end{eqnarray}
which does not depend any more on $\dot{\xi}$ and $\dot{a}$.

We now turn to Eqs.\ (\ref{eq:equsmotaxi}). By adding and subtracting $\braket{\rho_i}{f'(\eta_0)}$ in these equations, and substituting $\braket{\rho_i}{\partial_t\Delta\eta}$ by its expression derived from Eq.\ (\ref{eq:dconstr}), one finds
\begin{eqnarray}
0&=&\braket{\rho_i}{\mathcal{F}[\eta_0+\Delta\eta]}
=\braket{\rho_i}{\mathcal{F}[\eta_0]}\nonumber\\
&&{}+\braket{\rho_i}{\sigma_\eta[\Delta\eta]}+\braket{\rho_i}{f'(\eta_0)-f'(\eta_0+\Delta\eta)}\nonumber\\
&&{}+C\left\{\braket{\partial_\xi\rho_i}{\Delta\eta}\dot{\xi}+\braket{\partial_a\rho_i}{\Delta\eta}\dot{a}\right\},
\end{eqnarray}
where the last two lines represent an $O(\Delta\eta)$ correction to the leading-order mean-field equations, Eqs.\ (\ref{eq:xiao}).
\section{Numerical method}
\label{sec:nummeth}
The numerical procedure employed to solve the complex-valued EoM (\ref{eq:eomzetam}) warrants a detailed explanation. Up to $t=0$, the dislocation is assumed to move initially with constant velocity $\vi$, and core width $\ai$. These values must consistently be related together and with the initially applied stress by Eqs.\  (\ref{eq:steadystate}). The dislocation moves non-uniformly at times $t>0$, due to a change of applied stress.

Motion is discretized as a series of velocity jumps.\cite{PILL07} Effects of velocity jumps on the self-force have been widely studied in the past. In Eq.\ (\ref{eq:eomzetam}), the position and core-width variables stand on the same footing and must therefore be of same order of regularity. Thus, any velocity jump must go along with a jump of the core-width \emph{variation rate}. This information makes discretization straightforward, and avoids the complications of the simultaneous velocity and core-width jumps considered in Ref.\ \onlinecite{PELL12}.

Let velocity jumps occur at discrete times $t_k=k \delta t$, with $k$ a positive integer, and $\delta t>0$ the time step. Let moreover $t_{-1}=-\infty$. The characteristic function of the time interval $I_k=(t_k,t_{k+1})$ for $k\geq -1$ is
\begin{equation}
\theta_k(t)=\theta(t-t_k)-\theta(t-t_{k+1}),
\end{equation}
with $\theta_{-1}(t)=\theta(-t)$. The prescribed initial velocity is
\begin{equation}
\dot{\zeta}_{-1}\equiv \vi.
\end{equation}
The coordinate at $t=t_0=0$ is
\begin{equation}
\zeta_0\equiv\xi_0+(\ii/2)\ai,
\end{equation}
where $\xi_0$ is the reference position at which accelerated motion begins; and $\ai=a(v_i)$ by (\ref{eq:av}). Then $\zeta(t)=\zeta_0+\vi t$ for $t<0$, and the piecewise-constant complex velocity reads
\begin{equation}
\dot{\zeta}(t)=\sum_{k\geq -1}\dot{\zeta}_k\,\theta_k(t),
\end{equation}
where the $\dot{\zeta}_k$ must be determined for $k\geq 0$. Likewise, the imposed time-dependent force in (\ref{eq:eomzetam}) is sampled at intermediate times $t_{n+1/2}$ as
\begin{equation}
\label{eq:applf}
G_a^{(n)}\equiv(t_{n+1}-t_n)^{-1}\int_{t_n}^{t_{n+1}}\dd t\,G_a(t),
\end{equation}
where
\begin{equation}
G_a(t)\equiv -\ii b \sigth g\left(\frac{\sigma_a(t)}{\sigma_{\rm th}}\right)=-2\ii\frac{w_0}{d}g\left(\frac{\sigma_a(t)}{\sigma_{\rm th}}\right).
\end{equation}
Taking $\vi\not=0$ requires by (\ref{eq:steady}) $F_\alpha(\vi)$ to be equal to $G_a^{(-1)}$, namely, the constant force force applied before non-uniform motion.

Let, for positive times, $n=[t/\delta t]$ be the integer such that $t\in I_n$ (brackets denote the integer part). Positions at times $t=t_{n}$ are introduced as
\begin{equation}
\zeta_n=\zeta_0+\delta t\sum_{k=0}^{n-1} \dot{\zeta}_k\qquad (n\geq 0)
\end{equation}
where the sum is zero if $n=0$. Thus,
\begin{equation}
\label{eq:zetat}
\zeta(t)=\zeta_{n}+\dot{\zeta}_n (t-t_{n})\qquad (t>0).
\end{equation}
The adopted discretization requires us to compute the self-force (\ref{eq:sfzeta}) at time $t=t_{n+\frac{1}{2}}=t_n+\delta t/2$.
For $\tau\in I_k$, we define the following quantities:
\begin{equation}
\Delta\zeta^n_k=\zeta_{n}+\dot{\zeta}_n\frac{\delta t}{2}-\left\{
\begin{array}{l}
\zeta_{0}^*+\dot{\zeta}_k^*(n+1/2)\delta t,\\
\hspace{2.8cm}\text{if } k=-1,\\
\zeta_{k}^*+\dot{\zeta}_k^*(n-k+1/2)\delta t,\\
\hspace{2.8cm}\text{if } 0\leq k\leq n.
\end{array}
\right.
\end{equation}
We notice for further use that
\begin{equation}
\label{eq:dznn}
\Delta\zeta^n_n=2\ii\Im[\zeta_n+(\delta t/2)\dot{\zeta}_n].
\end{equation}
From (\ref{eq:vbar}) and (\ref{eq:zetat}) follows that
\begin{subequations}
\begin{eqnarray}
\label{eq:vbardisc}
\overline{v}(t_{n+\frac{1}{2}},\tau)&=&\dot{\zeta}_k^*+\frac{\Delta\zeta^n_k}{\Delta t},\\
\label{eq:dvbardisc}
\frac{\dd\overline{v}}{\dd \tau}(t_{n+\frac{1}{2}},\tau)&=&\frac{\Delta\zeta^n_k}{\Delta t^2},
\end{eqnarray}
\end{subequations}
where now $\Delta t=t_{n+\frac{1}{2}}-\tau$. With (\ref{eq:sfzeta}) and (\ref{eq:zetat}), the self-force at time $t_{n+1/2}$ is written as
\begin{eqnarray}
F^{(n)}_\zeta&=&2\left[\sum_{k=-1}^{n-1}\int_{t_k}^{t_{k+1}}+\int_{t_n}^{t_{n+\frac{1}{2}}^-}\right]\frac{\dd\tau}{\Delta t}m(\overline v)\frac{\dd\overline{v}}{\dd\tau}\nonumber\\
\label{eq:sfdisc}
&&{}+2\ii\frac{w_0}{\cS}\frac{\dot{\zeta}_n^*}{\Delta\zeta^n_n},
\end{eqnarray}
where the last (``local'') term has been written by appealing to (\ref{eq:dznn}). Since velocity is constant over each time interval, integrals can be carried out as in Sec.\ \ref{sec:steady}. Using (\ref{eq:vbardisc}) and (\ref{eq:dvbardisc}), one obtains
\begin{equation}
\label{eq:lint}
\int_{t_k}^{t_{k+1}}\frac{\dd\tau\,m(\overline v)}{t_{n+\frac{1}{2}}-\tau}\frac{\dd\overline{v}}{\dd\tau}=\frac{\Delta W^n_k-\dot{\zeta}^*_k \Delta p^n_k}{\Delta\zeta^n_k}
\end{equation}
where we introduce the following intermediate quantities for $-1\leq k\leq n$:
\begin{subequations}
\begin{eqnarray}
\label{eq:wnk}
\Delta W^n_k&=&
\left\{
\begin{array}{rl}
W(\overline{v}^n_{k+})-W(\overline{v}_{k}^{n})&\text{if } k<n\\
-W(\overline{v}_{n}^{n})&\text{if } k=n,
\end{array}
\right.
\\
\label{eq:pnk}
\Delta p^n_k&=&
\left\{
\begin{array}{rl}
p(\overline{v}^n_{k+})-p(\overline{v}_{k}^{n})&\text{if } k<n\\
-p(\overline{v}_{n}^{n})&\text{if } k=n,
\end{array}
\right.
\\
\overline{v}_{k+}^{n}&=&\dot{\zeta}^*_k+\frac{\Delta\zeta^n_k}{(n-k-1/2)\delta t},
\\
\overline{v}_{k}^{n}&=&
\left\{
\begin{array}{ll}
\vi+\ii\,0^+&\quad\text{if } k=-1,\\
\dot{\zeta}^*_k+\frac{\Delta\zeta^n_k}{(n-k+1/2)\delta t}&\quad\text{if } 0\leq k\leq n,
\end{array}
\right.
\end{eqnarray}
\end{subequations}
(remark that $\overline{v}_{k+}^{n}\not=\overline{v}_{k+1}^{n}$). Since $W(\ii\infty)=0$ and $p(\ii\infty)=\ii w_0/\cS$, the rightmost integral in (\ref{eq:sfdisc}) reduces to
\begin{eqnarray}
&&\hspace{-3em}\int_{t_k}^{t_{n+\frac{1}{2}}^-}\dd\tau\frac{m(\overline v)}{t_{n+\frac{1}{2}}-\tau}\frac{\dd\overline{v}}{\dd\tau}
=\frac{
\begin{array}{rl}
{}&W(\ii\infty)-W(\overline{v}_{n}^n)\\
{}&\qquad{}-\dot{\zeta}^*_n [p(\ii\infty)-p(\overline{v}_{n}^n)]
\end{array}
}{\Delta\zeta^n_n}\nonumber\\
\label{eq:rint}
&=&\frac{\Delta W^n_n-\dot{\zeta}^*_n \Delta p^n_n}{\Delta\zeta^n_n}
-\ii\frac{w_0}{\cS}\frac{\dot{\zeta}^*_n}{\Delta\zeta^n_n},
\end{eqnarray}
Substituting expressions (\ref{eq:lint}) and (\ref{eq:rint}) into Eq.\ (\ref{eq:sfdisc}) yields the following discretized expression of the self-force:
\begin{equation}
\label{eq:sfdisc2}
F^{(n)}_\zeta=
2\sum_{k=-1}^{n}\frac{\Delta W^n_k-\dot{\zeta}^*_k \Delta p^n_k}{\Delta\zeta^n_k},
\end{equation}
in which the contribution of the ``local" loss term in $\sigma_\eta$ has canceled out, in agreement with a remark made in Sec.\ \ref{sec:dpe}. The term $k=-1$ in the sum is a representation of  $F^{<}_\zeta(t_{n+1/2})$ [Eq.\ (\ref{eq:finfexpr})]. Expression (\ref{eq:sfdisc2}) is quite remarkable, as it involves only the known energy and momentum functions $W(v)$ and $p(v)$.\cite{HIRT98,PELL12}

Including the phenomenological drag, of same form as the last term in (\ref{eq:sfdisc}), the discretized EoM at time $t_{n+1/2}$ finally reads
\begin{equation}
\label{eq:discr}
E_n(\dot{\zeta}_n,\dot{\zeta}_n^*)\equiv F_\zeta^{(n)}+2\ii\alpha\frac{w_0}{\cS}\frac{\dot{\zeta}_n^*}{\Delta\zeta_n^n}-G_a^{(n)}=0,
\end{equation}
where $F_\zeta^{(n)}$ is given by (\ref{eq:sfdisc2}), and $G_a^{(n)}$ is given by (\ref{eq:applf}). Given $\dot{\zeta}_{-1}=\vi$, and assuming that the velocities $\dot{\zeta}_k$ have been computed for $0\leq k\leq n-1$, each term of the sum (\ref{eq:sfdisc2}) depends on $\dot{\zeta}_n$ ---the unknown at time step $n$--- for which (\ref{eq:discr}) constitutes an implicit complex-valued equation.

This equation is solved by the Newton-Raphson method. Using the shorthand notation $z=\dot{\zeta}_n$, and since $E_n$ is a non-holomorphic function, iterations read
\begin{subequations}
\begin{align}
z^{(0)}&=\left\{
\begin{array}{cc}
\dot{\zeta}_{-1}&\text{ if } n=0\\
2 \dot{\zeta}_{n-1}-\dot{\zeta}_{n-2}&\text{ if } n\geq 1
\end{array}
\right.,\\
\label{eq:newt}
z^{(k+1)}&=z^{(k)}+\frac{(\partial E_n/\partial z^*)E_n^*
-(\partial E_n/\partial z)^* E_n}{\left|\partial E_n/\partial z\right|^2-\left|\partial E_n/\partial z^*\right|^2},
\end{align}
\end{subequations}
where $k\geq 0$ is the iteration counter, $z^{(0)}$ is an initial convenient guess, and $E_n$ and its derivatives are evaluated at $z^{(k)}$. Equation (\ref{eq:newt}) follows from elementary formulas\cite{CART95} of complex-variable function theory. The derivatives are readily obtained as sums involving the functions $W'(v)=v\,m(v)$ and $p'(v)=m(v)$.

In all cases examined, we found this algorithm stable and inexpensive. No drift in the velocity occurs when applying the algorithm to an initial steady state, under a consistent value of $\sigma_a$. Reasonably well converged results are obtained with the time step $\delta t=\tau_0/10$, where $\tau_0=d/\cS$.



\begin{thebibliography}{99}

\bibitem{GUMB99}
P.~Gumbsch and H.~Gao,
Science \textbf{283}, 965 (1999);
J.\ Comput.-Aided Mater.\ Design \textbf{6}, 137 (1999).

%
\bibitem{LI02}
Q.~Li and S.Q.~Shi,
Appl.\ Phys.\ Lett.\ \textbf{80}, 3069 (2002).

%
\bibitem{MORD03}
D.~Mordehai, Y.~Ashkenazy, I.~Kelson and G.~Makov,
Phys.\ Rev.\ B \textbf{67}, 024112 (2003).

%
\bibitem{VAND04}
J.A.Y.~Vandersall and B.D.~Wirth,
Philos. Mag. \textbf{84}, 3755 (2004).

\bibitem{OLMS05}
D.L.\ Olmsted, L.G.\ Hector Jr., W.A.\ Curtin, and R.J.\ Clifton,
Modelling Simul.\ Mater.\ Sci.\ Eng.\ \textbf{13}, 371 (2005).

%
\bibitem{MARI06}
J.~Marian and A.~Caro,
Phys.\ Rev.\  B \textbf{74}, 024113 (2006).

\bibitem{MORD06}
D.~Mordehai, I.~Kelson and G.~Makov,
Phys.\ Rev.\ B \textbf{74}, 184115 (2006).

\bibitem{TSUZ08}
H.~Tsuzuki, P.S.~Branicio and J.P.~Rino,
Appl.\ Phys.\ Lett.\ \textbf{92}, 191909 (2008);
Acta Mat.\ \textbf{57}, 1843 (2009).

\bibitem{JING08}
Z.~Jin, H.~Gao and P.~Gumbsch,
Phys.\ Rev.\ B \textbf{77}, 094303 (2008).

\bibitem{PILL06}
L.\ Pillon, C.\ Denoual, R.\ Madec and Y.-P.\ Pellegrini,
J.\ Physique IV Proceedings \textbf{134}, 49 (2006).

\bibitem{NOSE07}
V.\ Nosenko, S.\ Zhdanov, and G.\ Morfill,
Phys.\ Rev.\ Lett.\ \textbf{99}, 025002 (2007);
 V.\ Nosenko, G.E.\ Morfill and P.\ Rosakis,
 \textit{ibid.} \textbf{106}, 155002 (2011).

\bibitem{MORF09}
G.E. Morfill and A.V. Ivlev,
Rev. Mod. Phys. \textbf{81}, 1353 (2009).

\bibitem{WEER69b}
J. Weertman, in \emph{Physics of Strength and Plasticity}, edited by A.~Argon (MIT Press, Boston, 1969), p.\ 75.

\bibitem{WEER69a}
J.~Weertman,
in \emph{Mathematical Theory of Dislocations}, edited by  T.\ Mura (American Society of Mechanical Engineers, New York, 1969), p.\ 178.

\bibitem{ROSA01}
P.~Rosakis,
Phys.\ Rev.\ Lett.\ \textbf{86}, 95 (2001).

\bibitem{GURR13}
B.~Gurrutxaga-Lerma, D.S.~Balint, D.~Dini, D.E.~Eakins and A.P.~Sutton,
Proc.\ R.\ Soc.\ A \textbf{469}, 201330141 (2013).

\bibitem{NADG88}
E.~Nadgornyi, Prog.\ Mater.\ Sci.\ \textbf{31}, 1 (1988).

\bibitem{LUND8586}
F.~Lund, Phys.\ Rev.\ Lett.\ \textbf{54}, 14 (1985); Bull. Seism.\ Soc.\ Am.\ \textbf{76}, 1790 (1986).

\bibitem{BITZ05}
E.~Bitzek and P.~Gumbsch, Mat.\ Sci.\ Engrgr.\ A \textbf{400--401}, 40 (2005).

\bibitem{HIRT98}
J.P.~Hirth, H.H.~Zbib and J.~Lothe,
Modelling Simul.\ Mater.\ Sci.\ Eng.\ \textbf{6}, 165 (1998).

\bibitem{PILL07}
L.~Pillon, C.~Denoual and Y.-P.~Pellegrini,
Phys.\ Rev.\ B \textbf{76}, 224105 (2007).

\bibitem{ESHE53}
J.D.~Eshelby,
Phys.\ Rev.\ \textbf{90}, 248 (1953).

\bibitem{BELT68}
R.J. Beltz, T.L. Davis, K. Mal\'en, Phys. Stat. Sol. (b) \textbf{26}, 621 (1968).

\bibitem{PELL12}
Y.-P.~Pellegrini,
J. Mech.\ Phys.\ Solids.\ \textbf{60}, 227 (2012).

\bibitem{CLIF81}
R.J.~Clifton and X.~Markenscoff,
J.\ Mech.\ Phys.\ Solids \textbf{29}, 227 (1981);
X.~Markenscoff and S.~Huang,
\textit{ibid.} \textbf{56}, 2225 (2008).

\bibitem{ALSH71}
V.I.~Al'shits, V.L.~Indenbom and A.A.~Shtol'berg, Zh.\ Eksp.\ Teor.\ Fiz.\ \textbf{60}, 2308 (1971) [Sov.\ Phys.\ JETP \textbf{33}, 1240 (1971)].

\bibitem{WEER67b}
J.~Weertman,
J. Appl. Phys. \textbf{38}, 5293 (1967).

\bibitem{MARI04}
J.~Marian, Wei Cai, and V.V.~Bulatov,
Nature Materials \textbf{3}, 158 (2004).

\bibitem{VITE68}
V.\ Vitek, Phil.\ Mag.\ \textbf{18}, 773 (1968).

\bibitem{SCHO01}
G.\ Schoeck, Phil.\ Mag.\ A \textbf{81}, 1161 (2001).

\bibitem{PEIE40}
R.E.~Peierls, Proc.\ Phys.\ Soc.\ \textbf{52} 34 (1940).

\bibitem{NABA47}
F.R.N.~Nabarro, Proc.\ Phys.\ Soc.\ \textbf{59}, 256 (1947).

\bibitem{HIRT82}
J.P.~Hirth and J.~Lothe,
\emph{Theory of dislocations},
2nd ed.\ (Wiley, New York, 1982).

\bibitem{SCHO05}
G.~Schoeck,
Mat.\ Sci.\ Enrgr.\ A \textbf{400-401}, 7 (2005).

\bibitem{PELL10}
Y.-P.~Pellegrini,
Phys.\ Rev.\ B \textbf{81}, 024101 (2010).

\bibitem{PELL11}
Y.-P.~Pellegrini,
Phys.\ Rev.\ B \textbf{83}, 056102 (2011).

\bibitem{BONN95}
M.~Bonnet,
\emph{Boundary Integral Equation Methods for Solids and Fluids}
(Wiley, New York, 1995).

\bibitem{DENO04}
C.~Denoual,
Phys.\ Rev.\ B \textbf{70} 024106 (2004);
Comput.\ Methods.\ Appl.\ Mech.\ Eng.\ \textbf{196}, 1915 (2007).

\bibitem{NINO72}
T. Ninomiya,
J. Phys. Soc. Jpn. \textbf{33}, 921 (1972);
\textbf{33}, 1235 (1972).

\bibitem{TOMB75}
E.\ Tomboulis, Phys. Rev. D \textbf{12}, 1678 (1975).

\bibitem{BOES88}
R.~Boesch, P. Stancioff and C.R. Willis, Phys.\ Rev.\ B \textbf{38}, 6713 (1988).

\bibitem{BOES90}
R. Boesch  and C.R.~Willis, Phys.\ Rev.\ B \textbf{42}, 6371 (1990).

\bibitem{SCHN00}
H.J. Schnitzer, F.G. Mertens and A.R. Bishop, Physica D \textbf{141}, 261 (2000).

\bibitem{PEYR04}
M.~Peyrard, T.~Dauxois,
\emph{Physics of Solitons} (Cambridge University Press, Cambridge, 2004).

\bibitem{MEDI06}
R.\ Medina, J.\ Phys.\ A: Math.\ Gen.\ \textbf{39}, 3801 (2006); F.\ Rohrlich, Phys.\ Rev.\ E \textbf{77}, 046609 (2008), and references therein.

\bibitem{NOTE1}
For brevity, we use indifferently in this Section the loose denominations ``force'' or ``stress" for what is technically a ``resolved shear stress'', i.e., a suitable projection of the stress on the glide plane.\cite{HIRT82}

\bibitem{MILL98}
R.~Miller, R.~Phillips, G.~Beltz and M.~Ortiz,
J.\ Mech.\ Phys.\ Solids.\ \textbf{46}, 1845 (1998).

\bibitem{GILM68}
J.J.~Gilman, Phys.\ Rev.\ Lett.\ \textbf{20}, 157 (1968).

\bibitem{NOTE2}
As a functional of $\eta$, the above expression of $\sigma_\eta(x,t)$ is exactly the same as in the dynamic theory of self-healing cracks. However, in the latter theory $\eta(x,t)$ has finite support so that $\widetilde{\eta}$ is written $\eta$.\cite{RICE93COCH94,PERR95} This connection was unknown to the author at the time Refs.\ \onlinecite{PELL10,PELL11} were published.
\bibitem{NOTE3}
Indeed, the kernel $K$ can be derived from the two-dimensional Green function of the Navier equation for the material dis\-pla\-ce\-ment.\cite{PELL10} It therefore encompasses all relevant wave physics, Bremsstrahlung included.

\bibitem{RICE93COCH94}
J.R.~Rice, J.\ Geophys.\ Res. \textbf{98}, 9885 (1993);
A.~Cochard and R.~Madariaga, Pure Appl. Geophys. \textbf{142}, 419 (1994).

\bibitem{WEER67a}
J.~Weertman,
J. Appl. Phys. \textbf{38}, 2612 (1967).

\bibitem{PERR95}
G.~Perrin, J.R.~Rice and G.~Zheng,
J.\ Mech.\ Phys.\ Solids \textbf{43}, 1461 (1995).

\bibitem{FRAN49}
F.C.~Frank,
Proc.\ Phys.\ Soc.\ A \textbf{62}, 131 (1949).

\bibitem{NOTE4}
As in Ref.\ \onlinecite{PELL12}, energy-related steady-state functions such as quasimomentum $p(v)$, Lagrangian $L(v)$, mass $m(v)$, or energy $W(v)$, are \emph{always} used here without their logarithmic factor---irrelevant to the present work. To avoid introducing additional symbols for these prelogarithmic terms \cite{BELT68}, same notations as for the usual functions with logarithmic factor included\cite{HIRT98} are employed.

\bibitem{LANC86}
C.~Lanczos,
\emph{The Variational Principles of Mechanics},
4th ed.\ (Dover, Mineola, 1986).

\bibitem{TCHO93}
P.\ Tchofo Dinda and C.R. Willis, Physica D \textbf{70}, 217 (1993).

\bibitem{BRAU04}
O.M.~Braun, Yu.S.~Kivshar,
\emph{The Frenkel-Kontorova Model: Concepts, Methods and Applications}
(Springer, Berlin, 2004).

\bibitem{JACK75}
J.D.\ Jackson,
\emph{Classical Electrodynamics},
2nd ed.\ (Wiley, New York, 1975), p.\ 790.

\bibitem{MORS53BART89}
P.M.~Morse and H. Feshbach,
\emph{Methods of Theoretical Physics} (McGraw Hill, New York, 1953), Vol.\ 1, p.\ 842;
G.~Barton,
\emph{Elements of Green's Functions and Propagation --- Potentials, Diffusion and Waves} (Clarendon Press, Oxford, 1989).

\bibitem{ROOS01}
A.~Roos, J.Th.M.~De Hosson and E.~Van der Giessen,
Comput.~Mat.~Sci.~\textbf{20}, 19 (2001).

\bibitem{STRO00}
S.H.\ Strogatz,
\emph{Nonlinear Dynamics and Chaos}
(Perseus, Cambridge, MA, 2000), p.\ 69.

\bibitem{MARE96}
T. Erneux and P. Mandel,
SIAM J.\ Appl.\ Math.\ \textbf{46}, 1 (1986);
G.J.M.~Maree,
\textit{ibid.} \textbf{56}, 889 (1996);
N.~Berglund and H.~Kunz,
J.\ Phys.\ A: Math.\ Gen.\ \textbf{32}, 15 (1999);
S.M.\ Baer and E.M.\ Gaekel,
Phys.\ Rev.\ E \textbf{78}, 036205 (2008).

\bibitem{ZHEN12}
Y.G.\ Zheng and Z.H. Wang,
Commun.\ Nonlinear Sci.\ Numer.\ Simulat.\ \textbf{17}, 3999 (2012), and references therein.

\bibitem{NEIS87}
A.I.~Neishtadt,
Differentsyal'nye Uravneniya \textbf{23}, 2060 (1987)
[Diff.\ Equations \textbf{23}, 1385 (1987)];
\textbf{24}, 226 (1988)
[\textbf{24}, 171 (1988)].

\bibitem{FINN84}
M.W.~Finnis and J.E.~Sinclair,
Philos.\ Mag.\ A \textbf{50}, 45 (1984);
G.\ Ackland and R.\ Thetford,
Philos.\ Mag.\ A \textbf{56}, 15 (1987).

\bibitem{KOIZ01}
H.~Koizumi, H.O.K.~Kirchner and T.~Suzuki,
Mat.\ Sci.\ Eng.\ A \textbf{309--310}, 117 (2001);
Phys.\ Rev.\ B \textbf{65}, 214104 (2002).

\bibitem{ESHE56}
J.D.~Eshelby,
Proc. Phys. Soc. B \textbf{69}, 1013 (1956).

\bibitem{GILB11}
M.R.~Gilbert, S.~Queyreau and J.~Marian,
Phys.\ Rev.\ B \textbf{84}, 174103 (2011).

\bibitem{CART95}
H.~Cartan,
\emph{Elementary Theory of Analytic Functions of One or Several Complex Variables}
(Dover, New York, 1995), Chap.\ II.

\end{thebibliography}
\end{document}